\documentclass[a4paper,11pt]{article}
\pdfoutput=1

\usepackage[a4paper,vmargin={30mm,30mm},hmargin={30mm,30mm}]{geometry}

\usepackage{lmodern}
\usepackage[T1]{fontenc}
\usepackage[utf8]{inputenc}

\usepackage[inline]{enumitem}

\usepackage{amsfonts}
\usepackage{amstext}
\usepackage{amsmath}
\usepackage{mathtools}
\usepackage{braket}

\usepackage{arydshln}
\usepackage[margin=\parindent,labelfont=bf]{caption}

\usepackage{pbox}

\usepackage[usenames,dvipsnames]{xcolor}

\usepackage{tikz}
\usetikzlibrary{decorations.markings}
\usetikzlibrary{arrows}

\numberwithin{equation}{section}
\numberwithin{figure}{section}

\usepackage[
colorlinks=true,
linkcolor=Red,
citecolor=Red,
urlcolor=Red,
unicode=true,
hypertexnames=false,
linktocpage=true,
bookmarksdepth=2
]{hyperref}
\hypersetup{
pdfauthor={
  Asger C. Ipsen, 
  Matthias Staudacher,
  Leonard Zippelius
}, 
pdftitle={
  title
   }
}

\usepackage{cite}
\usepackage[parsep]{collref}

\usepackage[yyyymmdd]{datetime}
\usepackage{currfile}

\usepackage[backgroundcolor=red!10!white,bordercolor=red,linecolor=red]{todonotes}

\DeclareMathOperator{\mul}{mul}
\newcommand{\lett}{A}
\newcommand{\flav}{\mathfrak{F}}
\newcommand{\noflav}{\emptyset}

\DeclareMathOperator{\tr}{tr}

\begin{document} 

\baselineskip 5mm

\begin{titlepage}

  \hfill
  \pbox{5cm}{
    \texttt{HU-Mathematik-2018-11}\\
    \texttt{HU-EP-18/39}
  }
  
  \vspace{2\baselineskip}

  \begin{center}

    \textbf{\LARGE \mathversion{bold}
      The One-Loop Spectral Problem of Strongly Twisted $\mathcal{N}$=4 Super Yang-Mills Theory\\
    }

    \vspace{2\baselineskip}

    Asger C. Ipsen,  Matthias Staudacher, and Leonard Zippelius
 
    \vspace{2\baselineskip}

    \textit{
     Institut für Mathematik und Institut für Physik, Humboldt-Universität zu Berlin,\\
      IRIS-Adlershof, Zum Großen Windkanal 6, 12489 Berlin, Germany\\
      \vspace{0.5\baselineskip}
    }

    \vspace{2\baselineskip}
    
     \texttt{
      \{acipsen,staudacher,lzippelius\}@physik.hu-berlin.de
    }

    \vspace{2\baselineskip}

    \textbf{Abstract}
    
  \end{center}

\noindent
We investigate the one-loop spectral problem of $\gamma$-twisted,        
planar $\mathcal{N}$=4 Super Yang-Mills theory in the double-scaling limit of infinite, imaginary twist angle and vanishing Yang-Mills coupling constant. This non-unitary model has recently been argued to be a simpler version of full-fledged planar $\mathcal{N}$=4 SYM, while preserving the latter model's conformality and integrability. We are able to derive for a number of sectors one-loop Bethe equations that allow finding anomalous dimensions for various subsets of diagonalizable operators. However, the non-unitarity of these deformed models results in a large number of non-diagonalizable operators, whose mixing is described by a very complicated structure of non-diagonalizable Jordan blocks of arbitrarily large size and with a priori unknown generalized eigenvalues. The description of these blocks by methods of integrability remains unknown.
\end{titlepage}

{\noindent}\hrulefill

\setcounter{tocdepth}{1}
\tableofcontents{}

\vspace{1\baselineskip}
{\noindent}\hrulefill
\vspace{1\baselineskip}


\section{\mathversion{bold}Introduction to Strongly Twisted \texorpdfstring{$\mathcal{N}$}{N}=4 Super Yang-Mills}
\label{sec:strongly twisted}

The discovery and exploitation of the integrability of planar $\mathcal{N}$=4 Super Yang-Mills theory (SYM) has been a huge success story. This was already the case when the overview collection \cite{Beisert:2010jr} appeared eight years ago. Since then the scope of planar integrability as concerns an ever increasing number of exactly computable quantities in $\mathcal{N}$=4 SYM and a small number of further, related integrable field theory models has been steadily increasing. An updated overview would certainly be warranted. On the other hand, vexingly, a convincing explanation on {\it why} these models are integrable at finite values of their coupling constants has so far not been discovered. Given this situation, a very interesting suggestion was made in \cite{Gurdogan:2015csr,Sieg:2016vap,Caetano:2016ydc,Chicherin:2017cns,Gromov:2017cja,Chicherin:2017frs,Grabner:2017pgm,Kazakov:2018hrh, Gromov:2018hut}. A certain non-unitary deformation of planar $\mathcal{N}$=4 SYM leads in a double-scaling limit to a decoupling of all gauge fields, a destruction of supersymmetry, and a vast simplification of the Feynman diagrammatics, while retaining integrability. It was then argued that, due to their perceived simplicity, a complete understanding of the integrability of these deformed models could be reached, with the final goal of feeding these insights back into a possible explanation of the integrability of the mother theory, $\mathcal{N}$=4 SYM. And indeed, the deformed models allow for a large number of exact computations that do appear simpler as compared to the undeformed case. On the other hand, oddly, the two showcase instances where integrability {\it can} be rigorously proved in $\mathcal{N}$=4 SYM are obscured in the double-scaled deformed models. At strong coupling, the interpretation in terms of an integrable string sigma model is missing, see, however, \cite{Basso:2018agi}. And at weak coupling, the a priori much simplified situation is also somewhat fuzzy. On the one hand, interesting connections between the Feynman diagrams of the double-scaled models and Yangian invariance have been made \cite{Chicherin:2017cns,Chicherin:2017frs}, integrability allowed to formulate a conjecture for an exact series of generalized ladder graphs \cite{Basso:2017jwq}, and a connection to an integrable $\mathfrak{su}(2,2)$ Heisenberg chain was made \cite{Gromov:2017cja}. On the other hand, we noticed that the integrable one-loop spin chain interpretation of the spectral problem of these models has not yet been properly exhibited. The purpose of this article is to begin a serious investigation of this issue. To anticipate: We find the one-loop spectral problem to be highly intricate, quite different from $\mathcal{N}$=4 SYM, and certainly unsolved.

For the rest of this introductory chapter, we collect a few pertinent facts on and notations for the strongly twisted cousins of $\mathcal{N}$=4 SYM that we investigate in this paper, focusing on the points important for the investigation of the one-loop spectral problem. We refer the reader to the original papers \cite{Gurdogan:2015csr,Sieg:2016vap,Caetano:2016ydc,Chicherin:2017cns,Gromov:2017cja,Chicherin:2017frs,Grabner:2017pgm,Kazakov:2018hrh, Gromov:2018hut} for motivations, derivations, and, above all, many more explanations and details.
For a generalization to dimensions other than four, albeit with non-local propagators, see \cite{Kazakov:2018qez,Derkachov:2018rot}.
The prime example of a four-dimensional integrable interacting quantum field theory, planar $\mathcal{N}$=4 SYM (see \cite{Beisert:2010jr} for some introductory material), allows for a deformation by three parameters $\gamma_i$ that appears to retain the integrability of the theory \cite{Lunin:2005jy,Frolov:2005dj}.
The resulting theory is called planar $\gamma$-twisted $\mathcal{N}=4$ SYM.
However, it had been demonstrated earlier in \cite{Fokken:2013aea} that, in contrast to undeformed $\mathcal{N}=4$ SYM, the $\gamma$-twisted theory is no longer conformally invariant, not even in the planar limit, due to running double-trace operator couplings. It was suggested by the authors of \cite{Gurdogan:2015csr} that this problem can nevertheless be circumvented in the planar model by restricting the attention to composite operators containing at least three fields. This allowed them to define an interesting double-scaling limit of the twisted models. Defining the squared planar gauge theory coupling constant as $g^2=\frac{\lambda}{16 \pi^2}$, where $\lambda$ is the `t Hooft coupling, they suggested to take $g\rightarrow 0$, while some or all of the twisting parameters $q_j=e^{-i \gamma_j /2}\rightarrow \infty$, such that the products $g\, q_j=\xi_j$ are held fixed. In this paper, we call these double-scaled, twisted models simply ``strongly twisted models''. Furthermore, we will mostly focus on two special cases. In the first one, all three $\xi_i$ are equal, i.e. $\xi_j=\xi$, and the model will be termed for ``historic reasons'' ($\gamma_i =\beta$) strongly $\beta$-twisted (s$\beta$t). In the second case $\xi_3=\xi$ while $\xi_{1,2}=0$. Here the model is often referred to as the fishnet theory (FN), as the resulting Feynman diagrams are reminiscent of a fishing net. In fact, the name was coined in an early paper by A.~Zamolodchikov \cite{Zamolodchikov:1980mb}, where the ``integrability'', in the sense of the validity of the star-triangle relation, for this class of diagrams was noticed. The two special cases result in the following two Lagrangian densities, respectively:
\begin{align}
\mathcal{L}_\text{int}^{\rm s\beta{\rm t}}&=N\xi^2\tr( \phi_1^\dagger \phi_2^\dagger \phi_1 \phi_2  +   \phi_3^\dagger \phi_1^\dagger \phi_3 \phi_1  +  \phi_2^\dagger \phi_3^\dagger \phi_2 \phi_3)  +iN\xi \tr( \psi_3\phi_1\psi_2+ \bar\psi_3 \phi_1^\dagger \bar\psi_2 + \text{cyclic} ), \label{eq:Lagrangianbeta}\\
\mathcal{L}_\text{int}^{\rm{FN}}&=N \xi^2 \tr( \phi_1^\dagger \phi_2^\dagger \phi_1 \phi_2), \label{eq:Lagrangianfishnet}
\end{align}
where by ``cyclic'' we mean cyclic permutations of the three indices, and we do not show the standard kinetic terms of complex bosonic and fermionic fields, but only the interacting part of the Lagrangian. 
It was shown in \cite{Sieg:2016vap} that the theories defined above, just like their unscaled counterparts \cite{Fokken:2013aea} (see above), are not perturbatively complete, and specific double-trace counter terms have to be added to the Lagrangian. It was then argued in \cite{Sieg:2016vap} and \cite{Grabner:2017pgm} that the couplings of these interaction terms may be fixed, such that the resulting theory becomes indeed conformally invariant in the planar limit, see also \cite{Mamroud:2017uyz}. Thus the local composite operators of these models should transform covariantly under the action of the dilatation operator of the conformal algebra. It is this dilatation operator of the strongly $\beta$-twisted model and the fishnet model in the planar limit that we investigate in detail at the leading one-loop order of perturbation theory in this paper. There, the double-trace interaction terms can safely be ignored for operators containing three or more fields, so we have omitted them from \eqref{eq:Lagrangianbeta}, \eqref{eq:Lagrangianfishnet}. They do not contribute in the planar limit unless they cut the color structure of a Feynman graph into two disjoint pieces.

A quick look at the Lagrangians \eqref{eq:Lagrangianbeta} and \eqref{eq:Lagrangianfishnet} suffices to realize that the hermitian conjugates of all the terms are missing.
This renders these models non-unitary, with a number of problematic technical and conceptual consequences. In particular, it leads to a non-hermitian dilatation operator that is potentially non-diagonalizable.
On the positive side, the absence of hermitian conjugates implies a dramatic reduction in the number of Feynman diagrams for a given quantity at each order of perturbation theory.
After all, this was the rationale behind the hope that these theories might provide interesting toy models for understanding the origins and consequences of integrability in diagrammatically more complex planar four-dimensional quantum field theories such as $\mathcal{N}$=4 SYM.

\section{\mathversion{bold}Spin Chain of Strongly Twisted \texorpdfstring{$\mathcal{N}$}{N}=4 Super Yang-Mills}
\label{sec:spinchain}
In this chapter we describe the general aspects of the spin chain picture of the strongly twisted theories at one-loop order, in close analogy with the original, undeformed case \cite{Minahan:2010js}.
We start by introducing a few technical notions in order to be able to then quickly derive the one-loop dilatation operator.
We end the chapter by introducing the novel notion of eclectic spin chain states with zero anomalous dimension.

\subsection{\mathversion{bold}Letters, Flavors and Chiral Ordering}
\label{subsec:letters}

We are interested in the action of the one-loop dilatation operator on single-trace operators built from the fields of $\mathcal{N}=4$ SYM
\begin{equation}
  \lett \in \{\partial^k \phi_i,\partial^k \phi_i^\dagger,\partial^k\psi_j,\partial^k\overline{\psi_j},\partial^k\mathcal{F},
  \partial^k\overline{\mathcal F}\},
  \label{eq:letters}
\end{equation}
where $i\in \{1,2,3\}$, $j\in \{1,2,3,4\}$, and we have suppressed all spacetime indices.
We will refer to the fields in \eqref{eq:letters} as letters or single-site spin chain states interchangeably. The number of letters in a single-trace operator equals the number of sites of the chain and is called the length $L$. Two letters appearing in the trace next to each other will be referred to as neighboring. For a more detailed review of the relation between single-trace operators and spin chains see for example \cite{Minahan:2010js}.
In order to describe the form of the dilatation operator, it is useful to define a map $\flav$ from letters to a set
of `flavors' $F_\noflav$,
\begin{equation}
  F := \{1,2,3,\bar{1},\bar{2},\bar{3}\},\qquad
  F_\noflav := F\cup \{\noflav\},
  \label{eq:flavorsets}
\end{equation}
via the assignments
\begin{align}
  \flav(\partial^k\phi_c) &= c\, ,& \flav(\partial_k\phi_c^\dagger) &= \bar{c}\, ,
    &\flav(\partial^k\psi_c) &= c\, , & \flav(\partial^k\overline{\psi}_c) &= \bar{c}\, , \label{eq:flavor1} \\
  \flav(\partial^k\psi_4) &= \noflav\, , & \flav(\partial^k\overline{\psi}_4) &= \noflav\, ,
    &\flav(\partial^k\mathcal{F}) &= \noflav\, , &\flav(\partial^k\overline{\mathcal F}) &= \noflav\, ,\label{eq:flavor2}
\end{align}
where $c=1,2,3$.
We further define $a_\pm=a\pm 1 \text{ mod }3$ and $\overline{a}_\pm=\overline{a\pm 1} \text{ mod }3$ for $a\in\{1,2,3\}$ and say two letters $AB$ are in chiral order, if $\flav(B) \neq \noflav$, and either $\flav(A)=\flav(B)_+$ or $\overline{\flav(A)}=\flav(B)_-$, where it is understood that $\overline{\overline{a}}=a$.
We say two letters $AB$ are in anti-chiral order if $BA$ are in chiral order.

Finally, we define $P^-$ and $P^+$ as chiral respectively anti-chiral {\it projection operators} acting on neighboring letters. They annihilate improperly ordered pairs of nearest neighbor states while leaving properly ordered pairs of states invariant.

\subsection{\mathversion{bold}The Dilatation Operator of Strongly Twisted \texorpdfstring{$\mathcal{N}$}{N}=4}
\label{subsec:dilatation}
Equipped with the above definitions, we are ready to derive the one-loop dilatation operator of the strongly twisted models from the one of the unscaled twisted models.
The dilatation operator $\mathfrak{D}$ of the conformal algebra consists of a classical part $\mathfrak{D}_0$ and quantum corrections $\delta\mathfrak{D}$:
\begin{equation}
\mathfrak{D}=\mathfrak{D}_0+\delta\mathfrak{D}\, .
\label{eq:Dilatationsplit}
\end{equation}
$\mathfrak{D}_0$ is identical for $\mathcal{N}=4$ SYM \cite{Beisert:2003jj} and all its deformations.
In this paper, we are interested in the one-loop contribution to $\delta\mathfrak{D}$, which we identify with the Hamiltonian $H$ of a spin chain
\begin{equation}
\delta\mathfrak{D}=\xi^2 H + \mathcal{O}(\xi^4)\, ,
\label{eq:Hamiltonian}
\end{equation}
where $\xi$ is the coupling constant of the strongly twisted theories see section \ref{sec:strongly twisted}.
Furthermore, $H$ is the sum of local Hamiltonian densities acting on neighboring spin chain sites as
\begin{equation}
H=\sum_{n=1}^L \mathcal{H}_{n,n+1}\, ,
\label{eq:Hamiltoniandensity}
\end{equation} 
where the sum is over spin chain sites $n$, and $L$ is the number of letters of the spin chain. As we consider single-trace operators, we impose periodic boundary conditions: $\mathcal{H}_{n,n+1}=\mathcal{H}_{n,1}$.
The matrix elements of the one-loop dilatation operator density $\mathcal H$ of 
unscaled\footnote{Note that the Hamiltonian derived from the density in \eqref{eq:Dgamma} 
should be multiplied by a factor of $g^2$, not $\xi^2$, to give the one-loop
contribution to $\delta\mathfrak{D}$.}
$\gamma$-twisted $\mathcal{N}$=4 SYM acting on a pair of neighboring letters $\lett_n$, $\lett_{n+1}$, given in \eqref{eq:letters}, at sites $n$ respectively $n+1$ is for $L\geq3$ \cite{Fokken:2013mza,Fokken:2013aea}
\footnote{We flip the sign of the exponent so as to have a notation consistent with the diagrams in \cite{Caetano:2016ydc}.}
\begin{equation}
(\mathcal{H}^\gamma)_{\lett_n \lett_{n+1}}^{\lett'_n \lett'_{n+1}}=\exp\left(\frac{-i}{2}\left((\mathbf{q}_{\lett_n})^{T} C \mathbf{q}_{\lett_{n+1}}+(\mathbf{q}_{\lett'_{n+1}})^{T} C \mathbf{q}_{\lett'_n}\right)\right) (\mathcal{H}^{\mathcal{N}=4})_{\lett_n \lett_{n+1}}^{\lett'_n \lett'_{n+1}}\, .
\label{eq:Dgamma}
\end{equation}
Here $\lett_n$, $\lett_{n+1}$ are the initial letters and $\lett'_n$, $\lett'_{n+1}$ the final letters as regards the action of this operator, and $\mathcal{H}^{\mathcal{N}=4}$ is the complete one-loop dilatation operator of the undeformed mother theory.
An explicit expression for $\mathcal{H}^{\mathcal{N}=4}$ was obtained in \cite{Beisert:2003jj}, but note that
our normalization is different: 
$\mathcal{H}^{\mathcal{N}=4}_{\text{here}} = 2 \mathcal{H}^{\mathcal{N}=4}_{\text{\cite{Beisert:2003jj}}}$.
The twist matrix is given in the conventions of \cite{Fokken:2013aea} as
\begin{equation}
C=\left( \begin{array}{ccc}
0 & -\gamma_3 & \gamma_2 \\
\gamma_3 & 0 & -\gamma_1 \\
-\gamma_2 & \gamma_1 & 0
\end{array} \right),
\label{eq:twistmatrix}
\end{equation}
and the charges $\mathbf{q}_{\lett}$ are vectors whose three components $q^i_\lett$ are shown in table \ref{tab:charges}.

\begin{table}[ht]
\begin{center}
\begin{tabular}{c|cccc|c|ccc}
  $\lett$  & $\psi_1$   & $\psi_2$   & $\psi_3$   & $\psi_4$   & $\mathcal F, \overline{\mathcal F}$   & $\phi_1$   & $\phi_2$   & $\phi_3$   \\ \hline
  $q_\lett^1$ &$+\frac{1}{2}$        &$-\frac{1}{2}$        &$-\frac{1}{2}$           &$+\frac{1}{2}$          & 0 & 1 & 0 & 0 \\ 
 $q_\lett^2$ &$-\frac{1}{2}$ &$+\frac{1}{2}$ &$-\frac{1}{2}$ &$+\frac{1}{2}$ & 0 & 0 & 1 & 0 \\ 
 $q_\lett^3$ &$-\frac{1}{2}$ &$-\frac{1}{2}$ &$+\frac{1}{2}$ &$+\frac{1}{2}$ & 0 & 0 & 0 & 1 \\ 
\end{tabular}
\end{center}
\caption{The charge vectors of the different fields in $\gamma$-twisted $\mathcal{N}$=4 SYM, cf.~\eqref{eq:letters}. Conjugate fields have the opposite charges. Gauge fields do not carry and derivatives do not add charge.}
\label{tab:charges}
\end{table}

When we scale the dilatation operator we have to keep in mind that it comes with a factor of the squared coupling constant $g^2$, which has to be combined with a factor of $q^2$ from the two coefficients in \eqref{eq:Dgamma} to yield something finite in the limit, see chapter \ref{sec:strongly twisted}.
We see that in the $\beta$-twisted model each of the two exponentials in \eqref{eq:Dgamma} produces a factor of $q$ as required, if and only if the fields change from chiral into anti-chiral order under the action of $H$.
Thus, as a first result, we find that the complete one-loop dilatation operator density of the strongly $\beta$-twisted model \eqref{eq:Lagrangianbeta} can be written for $L\geq3$ as\footnote{The case $L=2$ requires separate attention due to the double-trace terms that needed to be added to render \eqref{eq:Lagrangianbeta} and \eqref{eq:Lagrangianfishnet} conformal in the planar limit; cf.~our brief discussion of this in chapter \ref{sec:strongly twisted}.}
\begin{equation}
\mathcal{H}_{n,n+1}^{\rm s\beta{\rm t}}=P^+_{n,n+1}\mathcal{H}_{n,n+1}^{\mathcal{N}=4}P^-_{n,n+1}\, ,
\label{eq:Dbetascaled}
\end{equation}
where the projection operators $P^\pm$ have been defined at the end of the preceding subsection \ref{subsec:letters}.
The same argument carries over for the complete dilatation operator density $\mathcal{H}^{{\rm FN}}$ of the much simpler fishnet theory \eqref{eq:Lagrangianfishnet}.
The difference between the models lies in the coefficient functions in \eqref{eq:Dgamma}.
In particular, the matrix \eqref{eq:twistmatrix} includes the twist parameters $\gamma_i$ and hence depends on how we scale the different parameters.
We find that $\mathcal{H}^{{\rm FN}}_{n,n+1}$ is identical to $\mathcal{H}^{\rm s\beta{\rm t}}_{n,n+1}$ in \eqref{eq:Dbetascaled} {\it unless} any of the four letters $\lett_n$, $\lett_{n+1}$, $\lett'_n$ or $\lett'_{n+1}$ from \eqref{eq:Dgamma} are fermions, $\phi_3$, or $\phi_3^\dagger$, in which case the matrix element is zero. We could write this with the help of an additional projection operator $P^{{\rm FN}}$ that projects out these fields. That is,
\begin{equation}
\mathcal{H}_{n,n+1}^{\rm FN}=P^{{\rm FN}}_{n,n+1} P^+_{n,n+1}\mathcal{H}_{n,n+1}^{\mathcal{N}=4}P^-_{n,n+1} P^{{\rm FN}}_{n,n+1}\, .\label{eq:Dfishnet}
\end{equation}
More explicitly, for fishnet states without derivatives we find that the only non-vanishing matrix elements are (we hope that it is obvious that here the indices on the scalar fields are flavor indices, and that the pairs correspond to fields sitting on neighboring sites)
\begin{equation} 
(\mathcal{H}^{FN})_{\phi_2\phi_1}^{\phi_1\phi_2}=-2 \hspace{0.5cm} (\mathcal{H}^{FN})_{\phi_2^\dagger\phi_1^\dagger}^{\phi_1^\dagger\phi_2^\dagger}=-2 \hspace{0.5cm} (\mathcal{H}^{FN})_{\phi_1^\dagger\phi_2}^{\phi_2\phi_1^\dagger}=-2 \hspace{0.5cm} (\mathcal{H}^{FN})_{\phi_1\phi_2^\dagger}^{\phi_2^\dagger\phi_1}=-2\, .
\label{eq:fishnetD1}
\end{equation}
Including derivatives, some additional combinatorial factors carry over from $H_{n,n+1}^{\mathcal{N}=4}$, e.g.,
\begin{equation}
\mathcal{H}_{n,n+1}^{FN}\binom{M}{k}\left(\partial_{1\dot{1}}^k\phi_2\right)_n \otimes \left(\partial_{1\dot{1}}^{M-k}\phi_1\right)_{n+1}=- \frac{2}{M+1}\sum_{l=0}^M\binom{M}{l}\left(\partial_{1\dot{1}}^l\phi_1\right)_n \otimes \left(\partial_{1\dot{1}}^{M-l}\phi_2\right)_{n+1}\, .
\label{eq:fishnetD2}
\end{equation}
This matrix element can be extracted, for example, from equation (B.13) of \cite{Beisert:2003jj}.

The above results suffice for the two special cases we consider in the paper: The fishnet model and the strongly $\beta$-twisted model. It is not difficult to extend our arguments to more general strongly $\gamma_i$-twisted models, whose one-loop dilatation operator we spell out in appendix \ref{app:Dilatation_gamma} for future use.

\subsection{\mathversion{bold}Eclectic Spin Chains and Nilpotency of the Dilatation Operator}
\label{sec:eclectic}
Prior to turning to specific sectors of the strongly twisted theories, we would like to explain the nilpotency properties of their dilatation operator. To this end, we find it convenient to introduce the notion of eclectic spin chains.
As we can see from \eqref{eq:Dbetascaled}, \eqref{eq:Dfishnet}, and the discussion around these equations, for a given choice of vacuum, the Hamiltonian forces some of the flavors clockwise and some of the flavors anticlockwise around the spin chain.
For the rest of this paper, we will take $\phi_1$ as the spin chain vacuum, unless stated otherwise.
Furthermore, we will work in a convention where $\phi_2$, $\psi_{2 \alpha}$, $\phi_3^\dagger$ and $\bar{\psi}_{3\dot{\alpha}}$ are right-movers, their conjugates are left-movers, and the remaining fields will be called non-dispersing excitations for reasons to become clear later.
The non-dispersing excitations split into two groups: derivatives, which can only follow other excitations around the spin chain, and the remaining non-dispersing excitations, which simply never move at all.

If we add left and right-movers to the vacuum, a sufficient number of applications of the Hamiltonian will cause the excitations to meet.
The excitations can however not reflect off each other since otherwise they would travel in the wrong direction after scattering.
There are sectors in which the excitations can also not pass through each other.
Instead, they act as impenetrable walls towards one another.
Then the Hamiltonian density acting on these two excitations in the given order is identically zero.
We say operators corresponding to these types of spin chains have ``eclectic'' field content.
Explicitly, eclectic field content is given, if at least one of the following three conditions is met.
The operator contains fields of the three flavors $\{1,2,3\}$, or it contains conjugate fields of the three flavors $\{\overline{1},\overline{2},\overline{3}\}$, or it contains fields and conjugate fields of the same flavor\footnote{The above argument fails, if we {\it only} have fields and conjugate fields of the same flavor. However, then $H=0$. We take this special case to lie within our definition of eclectic field content.} $\{a,\overline{a}\}$.
The fermionic or bosonic nature of the excitations is irrelevant in our definition of eclectic field content, as are excitations of derivative type.
For eclectic chains, acting with the Hamiltonian a sufficient number of times will push all excitations against each other and then annihilate the spin chain state.
Thus the Hamiltonian is nilpotent; a rigorous proof of this statement can be found in appendix \ref{app:nilpotency-proof}.
We conclude that a large part of the complete one-loop dilatation operator of the strongly twisted theories has generalized eigenvalue zero.
Although the sizes of the Jordan blocks are not fixed by this argument\footnote{The size of the Jordan blocks is however bounded above for a given $L$ and given excitations, as can be seen directly from the proof in appendix \ref{app:nilpotency-proof}.}, we will focus on non-eclectic sectors for the rest of the paper. The reason is that we currently do not know if and how integrability may be used to determine the size of these blocks.

As an example consider the spin chain consisting of $L-2$ $\phi_1s$, one $\phi_2$ and one $\phi_3$.
A basis for this spin chain is:
\begin{equation}
\ket{l}=(-4)^l\ket{\phi_3 \phi_1^l \phi_2 \phi_1^{L-2-l}}-2\delta_{l,1} \ket{\phi_3 \phi_1^{L-2} \phi_2} \quad \text{for}\,\,\, 0\leq l \leq L-2\, .
\end{equation}
The Hamiltonian acts on this basis as: 
\begin{equation}
H\ket{l}=\ket{l+1},\, \text{when}\,\,\, l < L-2\qquad \text{and}\qquad H\ket{L-2}=0\, .
\end{equation}
Hence the Hamiltonian is just one Jordan Block of size $L-1$ and generalized eigenvalue zero, thus $H^{L-1}=0$ on this subspace.

Finally, we would like to relate the above findings to the results of \cite{Caetano:2016ydc}.
In this paper, the $\{\phi_1,\phi_2,\phi_3\}$ sector is discussed, and its asymptotic Bethe ansatz (ABA) equations are derived.
In our conventions, this would correspond to eclectic field content.
Hence, the one-loop limit of these equations should naively only yield zero-energy-states.
In fact, for general excitations, a Bethe ansatz should not work in this sector, since eigenstates in eclectic spin chains are not of the global form given by the Bethe ansatz.
However, we suspect that this apparent contradiction is merely due to a slight notational inconsistency in \cite{Caetano:2016ydc}.
While the limit, which the authors give in the introduction, is the same as ours, the twisted ABA equations they use in their appendix C differ by a replacement\footnote{The replacement is indeed only a convention. The theories before and after the replacement are identical up to a renaming of fields.} of $q_1\rightarrow q_1^{-1}$ and $q_2\rightarrow q_2^{-1}$. This changes the chirality of some of the vertices and the $\{\phi_1,\phi_2,\phi_3\}$ sector in this convention should be equivalent to a $\{\phi_1^\dagger,\phi_2^\dagger,\phi_3\}$ sector in our convention. 
Luckily, this sector is indeed not eclectic.
We discuss the one-loop Bethe equations of an equivalent sector, namely $\{\phi_1,\phi_2,\phi_3^\dagger\}$, in section \ref{sec:su3}.
The one-loop limit of the equations of \cite{Caetano:2016ydc} is difficult to compare directly to our results since we work in different gradings.
However, as a consistency check, it is possible to write our equations presented in section \ref{sec:su23} in the ABA grading and restrict to allow only specific excitation numbers to match the equations.
We found complete agreement in this case.

\section{\mathversion{bold}Fishnet Theory}
In this chapter, we investigate the non-eclectic sectors of the fishnet theory \eqref{eq:Lagrangianfishnet}.
Like all the strongly twisted theories this model is non-unitary and thus has a non-hermitian dilatation operator.
One therefore does not expect $H$ to be diagonalizable. And indeed, as explained in section \ref{sec:eclectic}, it turns out to be non-diagonalizable.
However, there are still proper eigenstates of $H$ and corresponding eigenvalues.
In this chapter, we show how to find these eigenvalues.
We propose four different methods for doing so: i) explicit construction of creation and annihilation operators, ii) an Algebraic Bethe Ansatz \cite{Faddeev:1996iy} from a strongly twisted R-matrix, iii) a Coordinate Bethe Ansatz \cite{Yang:1967bm} and iv) identification of the correct limit of the Beisert-Roiban Bethe equations for finite twists \cite{Beisert:2005if}. For a comparison of ii) and iii) as well as a discussion for finite twist see e.g.~\cite{Staudacher:2010jz}.
The technical details of the last method iv) are however deferred to the next chapter.
Not all these methods work in all cases, but whenever several of these methods are available we find perfect agreement between them.
For the broken $\mathfrak{su}(2)$ sector we can apply all four methods, while for the sector including one derivative only iii) and iv) are available and for the full fishnet model including any of the two scalars and any derivatives only iv) has yielded any results, although it would be interesting, if iii) were applicable.
Some emerging issues, features and open questions will be discussed at the end of this chapter.

\subsection{\mathversion{bold}The Broken \texorpdfstring{$\mathfrak{su}(2)$}{} Sector}
\label{subsec:xy}
The simplest non-trivial sector\footnote{Of course there are several equivalent discrete copies of this sector, e.g.~chains with only $\phi_1$, $\phi_2^\dagger$, etc.}
all the strongly twisted theories is the two-chiral-scalar-sector, where the letters of the spin chain are either $\phi_1$ or $\phi_2$, and neither the conjugate fields  $\phi_1^\dagger$, $\phi_2^\dagger$ nor any derivatives appear. In the original $\mathcal{N}$=4 model it corresponds to the $\mathfrak{su}(2)$ sector described by an integrable Heisenberg XXX spin chain.
Now the $\mathfrak{su}(2)$ symmetry is broken and the one-loop dilatation operator simply turns into a chiral permutation operator \cite{Caetano:2016ydc}, as seen from \eqref{eq:fishnetD1}.
It scans the spin chain until it finds two spins in chiral order and then exchanges them. Recall that we consider $\phi_1$ as a local vacuum state and $\phi_2$ as a local excitation.
In fact, $H$ is the Hamiltonian of a chiral XY-model, describing a free, chiral lattice fermion with a two-body S-matrix equalling -1. Being non-hermitian, we would a priori already expect the formation of non-diagonalizable Jordan\footnote{Camille Jordan, 1838-1922, Mathematician; Pascal Jordan, 1902-1980, Physicist. \label{jordan}} blocks. Interestingly, this is {\it not} yet the case. In fact, $H$ may be explicitly diagonalized by a Jordan\textsuperscript{\ref{jordan}}-Wigner transformation followed by a Fourier transform, just like the XY-model \cite{Lieb:1961fr}.
We refer the reader to this classic paper for the explicit construction. The eigenvalues $E$ of  the Hamiltonian $H$  are neatly and explicitly found to be
\begin{align}
E&=\sum_{j=1}^{K_4}\frac{-2}{\alpha_j}\, , \label{eq:cXYdispersion}\\
1&=\prod_{j=1}^{K_4}\alpha_j\, , \label{eq:cXYcyclicity}\\
\alpha_k^L&=(-1)^{K_4-1}\, , \label{eq:cXYquantization}
\end{align}
where the spin chain length $L$ is the total number of scalars $\phi_1$, $\phi_2$, and $K_4$ is the number of excitations $\phi_2$. The $\alpha_j=e^{i p_j}$ encode the lattice momenta $p_j$ of the excitations. They are determined by the free quantization laws \eqref{eq:cXYquantization} =``Bethe equations with S-matrix -1''. Unlike the case of true Bethe equations with a non-trivial S-matrix, the equations \eqref{eq:cXYquantization} may immediately be solved in terms of $L$-th roots of $(-1)$. Furthermore, again unlike the case of non-trivial Bethe equations, completeness of the eigenstates is easily demonstrated: At fixed $L$ and $K_4$ there are precisely (not necessarily cyclic) $\binom{L}{K_4}$ eigenstates, as expected. The projection to cyclic states, as required by the trace of the model's composite operators, is encoded in \eqref{eq:cXYcyclicity}. Finally, the dispersion law is read off from \eqref{eq:cXYdispersion}, clearly showing that there are only right-movers: $E \sim \sum_j e^{-i p_j}$. These equations possess a hidden symmetry\footnote{The XY-model, chiral or not, does not have a ``beyond the equator problem'' at $K_4 > \frac{L}{2}$, unlike the XXX Heisenberg spin chain: Given $L$, for $K_4$ all values with $0 \leq K_4 \leq L$ are allowed.}: The spectrum for $E$ is invariant under the replacement $K_4 \rightarrow L-K_4$, interchanging vacuum states $\phi_1$ and excitations $\phi_2$, thereby also interchanging right and left-movers. One simple example of eigenstates in the broken $\mathfrak{su}(2)$ sector are for $L=4$ and $K_4=2$ the operators $O=\pm \sqrt{2} \tr (\phi_1\phi_1\phi_2\phi_2)+\tr(\phi_1\phi_2\phi_1\phi_2)$, for which we find $E=\mp 2 \sqrt{2}$, in agreement with \cite{Caetano:2016ydc}.

\subsection{\mathversion{bold}Including Derivatives}
\label{sec:fishnetderivative}
As mentioned earlier, we only consider non-eclectic sectors, since the generalized eigenvalues corresponding to eclectic spin chain states are zero.
Thus, in order to extend beyond the two-scalar-sector within the fishnet theory, we can only add derivatives, since any additional scalar would produce eclectic field content and the Hamiltonian would become nilpotent.
For simplicity, we start by introducing derivatives of only one kind, say\footnote{We use Weyl notation for the derivatives, i.e.~$\partial_\mu \rightarrow \partial_{\alpha \dot{\alpha}}$.} $\partial_{1\dot{1}}$.
Individual spins are then taken out of $\{\partial_{1\dot{1}}^k\phi_1,\partial_{1\dot{1}}^k\phi_2\}$, and we drop the spinor indices on the derivatives in the following to avoid too many subscripts: 
$\partial_{1\dot{1}} \rightarrow \partial$.
In contrast to other excitations, we can have an arbitrary number of derivatives at each spin chain site.
As before, our aim is to diagonalize $H$, which we attempt to achieve by a coordinate Bethe ansatz.
However, as it turns out, the Hamiltonian is non-diagonalizable in this sector, and ultimately the Bethe ansatz is incomplete. Nevertheless, it does find a set of eigenvalues, as we will see in the next section.

Here, we do not describe the full machinery of the coordinate Bethe Ansatz, but instead refer the reader to the existing literature, e.g.~\cite{Staudacher:2004tk,Beisert:2005fw}.
Within the Bethe ansatz the individual excitations carry lattice momenta, and the eigenvalue of the Hamiltonian $H$, i.e., the energy $E$, is given in terms of these momenta in the dispersion law.
Acting on single-excitation states, we find that only the momentum of the $\phi_2$ excitations enters the dispersion law because a single derivative excitation cannot move by itself.
Thus, the set of the momenta distributed between the $\phi_2$ has to stay the same during scattering, and we conclude that the momenta cannot be exchanged between $\phi_2$ excitations and $\partial$ excitations.
This property simplifies the action of the S-matrix of spin chain excitations to be an exchange of the excitations of type $f_1$ and $f_2$ multiplied by a scalar function $S_{f_1,f_2}$, explicitly
\begin{align}
S|\phi_2 (p_1)\partial (p_2)\rangle=&S_{\phi_2,\partial}(p_1,p_2)|\partial (p_2)\phi_2 (p_1)\rangle=\frac{e^{ip_2}}{2-e^{-ip_2}} |\partial (p_2)\phi_2 (p_1)\rangle, \label{eqSphid}\\
S|\partial (p_1)\phi_2 (p_2)\rangle=&S_{\partial,\phi_2}(p_1,p_2)|\phi_2 (p_2)\partial (p_1)\rangle=\frac{2-e^{-ip_1}}{e^{ip_1}}|\phi_2 (p_2)\partial (p_1)\rangle,\\
S|\phi_2 (p_1)\phi_2 (p_2)\rangle=&S_{\phi_2,\phi_2}(p_1,p_2)|\phi_2 (p_2)\phi_2 (p_1)\rangle=-|\phi_2 (p_2)\phi_2 (p_1)\rangle,\\
S|\partial (p_1)\partial (p_2)\rangle=&S_{\partial,\partial}(p_1,p_2) |\partial (p_2)\partial (p_1)\rangle = -\frac{e^{i(p_1+p_2)}-2e^{ip_2}+1}{e^{i(p_1+p_2)}-2e^{ip_1}+1}|\partial (p_2)\partial (p_1)\rangle\, . \label{eqSdd}
\end{align}
The first three equations may be derived in the usual way by considering two-excitation states.
The scattering of derivatives requires to consider a spin chain with two $\partial$ excitations. However, since the derivatives cannot move by themselves they do not scatter unless we also add an additional $\phi_2$ as a transporter. To obtain the fourth equation, we considered a state with the three excitations: $\{\partial, \partial, \phi_2\}$.
As discussed above, if in a given ordering of excitations we assign the flavors to the momentum, the S-matrix preserves this assignment.
Put differently, if in one ordering of the excitations the particle with momentum $p_i$ has flavor $f_i$, it will have flavor $f_i$ in all orderings of the excitations.
This property allows us to make the following coordinate Bethe ansatz, using only a scalar function $S_\sigma$ for an eigenstate with a general number of excitations $M$
\begin{equation}
\ket{\Psi}=\sum_{n_1<n_2<...<n_M}\sum_{\sigma} S_{\sigma} (\{p_i\}) e^{i\sum_j p_{\sigma(j)}n_j} \ket{n_1,...,n_M;f_{\sigma(1)},...,f_{\sigma(M)}}+\text{local terms}, \label{eq:derivativeCBA}
\end{equation}
where we sum over all permutations $\sigma$ of M elements and $\ket{n_1,...,n_M;f_{\sigma(1)},...,f_{\sigma(M)}}$ is the state with excitations of type $f_{\sigma(i)}$ at spin chain sites $n_i$.
$S_\sigma$ is the product of S-matrices $S_{f_mf_n}$ in \eqref{eqSphid}-\eqref{eqSdd} such that the excitations are scattered into the order given by $\sigma$.
The Yang-Baxter equation guarantees that different decompositions of the permutation $\sigma$ agree.
Imposing periodic boundary conditions on \eqref{eq:derivativeCBA} leads to the Bethe equations
\begin{equation}
1=e^{ip_i L} \prod_{i \neq j} S_{f_i f_j}(p_i,p_j).
\label{eq:BAfishnetder}
\end{equation}
Remarkably, the diagonal form of the S-matrix has eliminated the standard nesting, which is usually seen in spin chains with several different excitations.
The Bethe equations \eqref{eq:BAfishnetder} then take the form 
\begin{align}
e^{ip_{\phi,k}L} &= (-1)^{K_R-1} \prod_{j=1}^{K_\partial}\frac{2-e^{-ip_{\partial,j}}}{e^{ip_{\partial,j}}}\, , \label{eq:BAfishnetderexplicit1}\\
e^{ip_{\partial,k}L} &= \left(\frac{e^{ip_{\partial,k}}}{2-e^{-ip_{\partial,k}}} \right)^{K_R} \prod_{j\neq k}^{K_\partial}-\frac{e^{i(p_{\partial,k}+p_{\partial,j})}-2e^{ip_{\partial,j}}+1}{e^{i(p_{\partial,j}+p_{\partial,k})}-2e^{ip_{\partial,k}}+1}\, ,\label{eq:BAfishnetderexplicit2}\\
E&=\sum_{j=1}^{K_R}\frac{-2}{e^{ip_{\phi,j}}}\, ,\label{eq:BAfishnetderexplicit3}\\
1&=\prod_{j=1}^{K_R}e^{ip_{\phi,j}} \prod_{j=1}^{K_\partial}e^{ip_{\partial,j}}\, ,\label{eq:BAfishnetderexplicit4}
\end{align}
where we have also added the dispersion formula and and the cyclicity relation. Here, $L$ is the number of scalars, $K_\partial$ and $K_R$ are the number of derivatives $\partial$ and right-movers $\phi_2$ respectively, which have momenta $p_\partial$ and $p_\phi$, and $E$ is the eigenvalue of $H$.
It is worth mentioning that the scattering of a scalar with a derivative only depends on the momentum of the derivative and not on the momentum of the scalar.
This is unusual for spin chains and simplifies the solution of the Bethe equations.
However, the equations presented here are no longer manifestly solvable as was the case for \eqref{eq:cXYquantization},\eqref{eq:cXYdispersion},\eqref{eq:cXYcyclicity} 
in the previous section.

The full fishnet model has eclectic field content, since it includes scalars and their conjugates.
The largest non-eclectic part of the model includes two scalars and four types of derivatives.
In principle a Bethe ansatz as described above should be possible. However, for this sector we shall restrict ourselves for simplicity to scaling the Beisert-Roiban equations, as will
be presented later in section \ref{sec:fishnet-scaling}.
For now we just mention that equations \eqref{eq:BAfishnetderexplicit1}, \eqref{eq:BAfishnetderexplicit3} and \eqref{eq:BAfishnetderexplicit4} will stay the same; only \eqref{eq:BAfishnetderexplicit2} will be replaced by a nested system of three other levels of Bethe equations.

\subsection{\mathversion{bold}Completeness of the Bethe Equations: Observations and Results}
\label{sec:completeness}
Whether the Bethe ansatz gives the entire spectrum of the Hamiltonian of a spin chain is an interesting and long-standing problem, see for example \cite{Hao:2013rza} and \cite{Hao:2013jqa} for some recent work and further references.
For the broken $\mathfrak{su}(2)$ sector, see section \ref{subsec:xy},  we mentioned already that the ``Bethe ansatz equations'' \eqref{eq:cXYquantization} for this spin chain (the chiral XY-model) yield the full spectrum of $2^L$ states. The subset of cyclic states that we need are then found from the condition of total zero momentum \eqref{eq:cXYcyclicity}. Jordan blocks are absent. This is no longer the case as soon as we leave this simplest sector of the fishnet model. Let us summarize our ``experimental'' findings, which all appear to be new:

\begin{enumerate}

\item Adding derivatives, we find that (even) the Hamiltonian of the fishnet model develops Jordan blocks. 
It is currently unclear, if, and if so, how one can find these with an ansatz of Bethe type.
However, see the encouraging paper \cite{Gainutdinov:2016pxy},  which contains a modification of the Bethe ansatz for an -- albeit different -- model that also possesses Jordan blocks.

\item Note that every Jordan block does include exactly one proper eigenstate.
We found experimentally, in the case of the fishnet model, that for small length and low excitation numbers these eigenstates are not of Bethe type either, unfortunately.
	
\item The Jordan blocks are not all of size\footnote{In contradistionction to the model in \cite{Gainutdinov:2016pxy}.} $2 \times 2$, they do become arbitrarily  large with increasing $L$.
We demonstrate this by an explicit example below in \eqref{eq:state-with-wall}. It is unclear to us how to determine these sizes, in general.

\item There appears to be an abundance of Jordan blocks once we include derivatives. Their systematics and counting is unclear to us.

\item The generalized eigenvalue of all Jordan blocks we found turned out to be $E=0$ in all instances. It is natural to conjecture that this is always the case. However, so far we could not yet find a proof of the latter.
\end{enumerate}

Clearly it would be interesting and important to prove these observations, especially the last one, and to establish a complete classification into eigenstates and Jordan cells of all states of the fishnet model. As a possible first step, we discovered certain ``wall-like structures'' that can be constructed out of a $\phi_2$ decorated with arbitrarily many derivatives.
The simplest wall is of the form 
\begin{equation}
\ket{\ldots \Big( (\partial\phi_2) \phi_1-\phi_2(\partial\phi_1) \Big) \ldots},
\label{eq:wall}
\end{equation}
where by $ \ldots $ we indicated the rest of the spin chain.
The Hamiltonian annihilates this part of the spin chain. In addition, this structure represents an impenetrable wall for any other $\phi_2$ traveling around the chain.
This is reminiscent of the situation we have for eclectic spin chains, and indeed we conclude that we have subspaces of states containing such walls where the Hamiltonian is again nilpotent.
As an explicit example, consider the state
\begin{equation}
  \ket{\psi} := \ket{(\partial\phi_2)\phi_1\phi_2\phi_1^{L-3}}
    -\ket{\phi_2(\partial\phi_1)\phi_2\phi_1^{L-3}} .
  \label{eq:state-with-wall}
\end{equation}
It is easy to check that $H^{L-3}\ket{\psi} \neq 0$, but $H^{L-2}\ket{\psi} = 0$.
This means that this operator is part of a Jordan block of size (at least) $L-2$.
A more detailed discussion can be found in appendix \ref{app:walls}.

We know that the Bethe ansatz does not even give all the eigenstates not belonging to Jordan blocks since we found the following counterexample. Let us consider a spin chain of general length $L$, with a single $\phi_2$ and a single $\partial_{1\dot{1}}$.
The Bethe equations for such a spin chain are given in \eqref{eq:BAfishnetderexplicit1} to \eqref{eq:BAfishnetderexplicit4}.
We can rewrite them in terms of a single polynomial equation for the energy $E$ as
\begin{equation}
2\left(\frac{E}{-2}\right)^{L-1}-\left(\frac{E}{-2}\right)^{L-2}-1=0\, . \label{eq:examplefishnetderivative}
\end{equation}
This equation has $L-1$ solutions, but the subspace is $L$ dimensional, therefore we immediately see that the Bethe equations are missing one eigenstate.
The missing eigenstate is exactly the wall we talked about in the last paragraph. In some physical (but not mathematical) sense this is a kind of $1 \times 1$ Jordan block, as the simplest member of the family of wall-containing states.
The spin chain also includes a conformal descendant, which is the solution with\footnote{We find the descendant using the Bethe equations, since we work on the level of momenta and not of the usual Bethe roots. The descendant still corresponds to a singular Bethe root $u$, but a finite $e^{ip}$.} $E=-2$. For large enough $L$ we can see that $|E/2|$ has to be close to one for the terms in \eqref{eq:examplefishnetderivative} to be of roughly the same size.
So for larger $L$ the derivative is less impactful, and the  spectrum (except for the single eigenvalue $E=0$) becomes approximately that of the spin chain without the derivative and without the cyclicity constraint imposed.\footnote{Indeed, it is easy
to see that the solutions to \eqref{eq:examplefishnetderivative} for large $L$ are $-E_n/2 = \omega_n\exp[-(L-1)^{-1}\log(2-1/\omega_n)+O(L^{-2})]$
with $\omega_n = \exp[2\pi i n/(L-1)]$, $n=0,1,\ldots,L-2$.}
This means for large $L$ the derivative acts only as a reference point to which the distance of the $\phi_2$ can be measured and not as a properly traveling excitation.

Before we mention a few open questions on the completeness of the Bethe equations, let us use the above example to see how the Bethe equations behave as one approaches the double scaling limit from the unscaled twisted theory.
Proceeding numerically, it is simplest to choose $L=3$.
In the unscaled theory the sector with a $\phi_2$ and a $\partial_{1\dot{1}}$ is not closed. There are additional states, where the excitations have combined to form a $\bar\psi_{3\dot{1}}$ and a $\psi_{41}$.
The eigenstates are given by two states including only $\phi_2$ and $\partial_{1\dot{1}}$ excitations as above, one eigenstate including only  a $\bar\psi_{3\dot{1}}$ and a $\psi_{41}$ excitation and two eigenstates consisting out of the wall with some additional terms including a $\bar\psi_{3\dot{1}}$ and a $\psi_{41}$ excitation.
In the strong twisting limit the $\bar\psi_{3\dot{1}}$ and $\psi_{41}$ decouple, hence we can look at the $\{ \phi_2\partial_{1\dot{1}} \}$ sector by itself.
The projections of the eigenstates containing the wall onto the $\{ \phi_2\partial_{1\dot{1}} \}$ sector are identical, but not of Bethe form.
The Bethe form only comes from the inclusion of the fermionic sector.
This explains both, why the Bethe ansatz does not find these, as well as how in general Bethe states can get lost in the double scaling limit, which allows for Jordan block formation.
The only other possibility we have found how Bethe states can be lost in the limit is by a collision of solutions. However, we have only observed this phenomenon in the $\beta$-twisted theory \eqref{eq:Lagrangianbeta}.

\subsection{\mathversion{bold}Completeness of the Bethe Equations: Open Questions}
Let us pause to discuss the observations from the last section a bit further. After all, they are the basis for our claim that the strongly twisted models are rather different from their ``mother theory'', i.e.~untwisted $\mathcal{N}=4$ SYM, and as such somewhat unsuitable for clarifying the origins of integrability of the latter.
So far we have only found Jordan blocks with generalized eigenvalue zero. Is this a general feature or are there Jordan blocks with non-zero generalized eigenvalues?
For eclectic spin chains, as well as wall-like structures of the last section, $E=0$ immediately follows. Put differently, are there other mechanism for Jordan block formation?
Or else, is there a basis in which the Hamiltonian may be manifestly written as the direct sum of a diagonalizable part and a wall subspace?

The above questions concern the spin chain and the spectrum of the dilatation operator as such.
It is also interesting to understand how the Bethe ansatz and the quantum inverse scattering method 
relate to these open problems. Can integrability at least find all non-zero eigenvalues? And can it quantitatively describe the Jordan cell formation of these integrable non-unitary model(s)?
Finally one might wonder whether one can combine the Bethe ansatz with a proper investigation of the walls to find the full spectrum of the non-eclectic part of the spin chain?\footnote{One might be tempted to pose this question as follows: Can the Bethe ansatz diagonalize the wall-free subspace? However, note that the wall-free subspace is not yet rigorously defined. While the wall subspace as such may be defined as in appendix \ref{app:walls}, to define an orthogonal complement we would have to introduce a suitable inner product, which appears unnatural to us.}

These issues are also present in the case of the more complicated $\beta$-twisted theory \eqref{eq:Lagrangianbeta}. Before turning to the latter, we will first expand our machinery.

\section{\mathversion{bold}Scaling Limit of the Twisted Bethe Equations}
\label{sec:scaling}
In order to prepare for the discussion of the spectrum of the strongly $\beta$-twisted theory \eqref{eq:Lagrangianbeta}, we require a way to obtain one-loop Bethe equations in more complicated sectors. Therefore we discuss in this chapter the scaling procedure on the level of the general twisted one-loop $\mathcal{N}=4$ Bethe equations, originally worked out in \cite{Beisert:2005if}. We refer to this paper and references therein for background information on these equations. In addition, more details on the notation are given in appendix \ref{app:twisted-bethe-equations}.

\subsection{\mathversion{bold}Basic Scaling}
\label{sec:basic-scaling}

The scaling of the momentum-carrying roots $u_{4,j}$, which encode the lattice momentum of the excitations, can be fixed by considering
the dispersion relation. To illustrate this, we take a plane wave of a single excitation $A$ 
on an infinite chain (with $\phi_1$ as the vacuum),
\begin{equation}
  |p\rangle = \sum_n e^{ipn} |A(n)\rangle ,
\end{equation}
where the sum is over spin chain sites n.
Acting with the $\beta$- or $\gamma_3$-twisted Hamiltonian  \eqref{eq:Hamiltoniandensity},
\eqref{eq:Dgamma},
with $q=e^{-i \beta / 2}$ or $q=e^{-i \gamma_3 /2}$ respectively we find, before taking the scaling limit,
\begin{equation}
  H_q |p\rangle  = E_q(p)|p\rangle := (4-2q^{2c}e^{-ip}-2q^{-2c}e^{ip})|p\rangle ,
\end{equation}
where $c$ is $+1$, $-1$ or $0$ for a right-moving, left-moving, or non-dispersing excitation, 
respectively.\footnote{For the $\gamma_3$-twist the fermions will have $c=\pm 1/2$, but, since these
  fields decouple completely in the strong twisting limit, we will ignore them.}
On the other hand, the dispersion law of the Beisert-Roiban equations is\cite{Beisert:2005if}
\begin{equation}
  E(u_4) = 2i\left[\frac{1}{u_4+i/2}-\frac{1}{u_4-i/2}\right], 
  \label{eq:u-dispersion}
\end{equation}
with no explicit dependence on $q$. It follows that the relation between $u_4$ and $p$ is
\begin{equation}
 e^{ip} = q^{2c} \frac{u_{4}+i/2}{u_{4}-i/2}\, .
\end{equation}
When taking the strong twisting limit $q \rightarrow \infty$ we are focusing on states where the spin chain momenta $p_j$ remain finite. This is conveniently implemented by the change of variables
\begin{equation}
  u_{4,j} \to \begin{cases}
            -i/2 - iq^{-2}\alpha_{4,j}\, ,      &j=1,\ldots,K_R\, ,\\
  			 +i/2 + iq^{-2}\alpha_{4,j-K_R}'\, , &j = K_R+1,\ldots,K_R+K_L\, ,\\
  			 \tilde{u}_{4,j-K_R-K_L}\, ,         &j = K_R+K_L+1,\ldots, K_4\, ,
          \end{cases} 
  \label{eq:ufour-scaling}
\end{equation}
where $K_R$ ($K_L$) is the number of right- (left)-movers.
For e.g.~a right-mover we then have the identification $e^{ip} = \alpha_4$ in the $q\to\infty$ 
limit.\footnote{The wave function corresponding to a given set of Bethe roots will, in general, 
  contain terms pairing the different types of excitations with the momenta in all possible ways.
  One would thus expect to find terms where, say, a right-mover is paired with a left moving momentum.
  For the wave function to be non-singular with the above scaling of $u_4$ it is necessary that the amplitude 
  of such terms is zero in the $q\to\infty$ limit (or that some kind of subtle cancellation occurs).}
The total energy becomes
\begin{equation}
  E = \lim_{q\to\infty} q^{-2}\sum_{j=1}^{K_4}E_q(u_{4,j}) 
    = -2\sum_{j=1}^{K_R}\frac{1}{\alpha_{4,j}}-2\sum_{j=1}^{K_L}\frac{1}{\alpha'_{4,j}}\, .
\end{equation}
This should be the eigenvalue of the strongly $\beta$-twisted or fishnet Hamiltonian with densities \eqref{eq:Dbetascaled} and \eqref{eq:Dfishnet}.
Before tackling the actual Bethe equations, let us deal with the zero-momentum=cyclicity constraint.
According to appendix \ref{app:twisted-bethe-equations} it reads
\begin{equation}
  q^{2K_R-2K_L}\prod_{j=1}^{K_4}\frac{u_{4,j}+i/2}{u_{4,j}-i/2} = 1\, .
  \label{eq:genmomconstr}
\end{equation}
Substituting our change of variables and taking the $q\to\infty$ limit we find
\begin{equation}
  \prod_{j=1}^{K_R}\alpha_{4,j}\prod_{j=1}^{K_L}\frac{1}{\alpha'_{4,j}}
     \prod_{j=1}^{K_4-K_R-K_L}\frac{\tilde{u}_{4,j}+i/2}{\tilde{u}_{4,j}-i/2} = 1\, .
\end{equation}
The correct scaling of the auxiliary roots is more subtle and is highly dependent on the
sector under consideration. In the next sections we will discuss several cases where we, to a large extent,
are able to check our ans{\"a}tze by cross-checking the resulting
Bethe equations against other methods.

\subsection{\mathversion{bold}Application to Fishnet Theory}
\label{sec:fishnet-scaling}
Here we treat the full chiral fishnet sector that we already briefly mentioned in section \ref{sec:fishnetderivative}. It consists of $\phi_1$ (vacuum), $\phi_2$ (excitation) and any of the four derivatives $\partial_\mu \sim \partial_{\alpha \dot \alpha}$ (four further excitations).
The twisted Bethe equations are given in appendix \ref{app:ABA-bethe-equations}, and this specific sector is obtained
by restricting auxiliary roots according to $K_1 = K_7 = K_3 - K_5 = 0$. We denote the number
of derivatives by $K_\partial = K_3 = K_5$, and the number of right-movers ($\phi_2$'s)
is $K_R = K_4 - K_\partial$.

Let us first consider a state of a single $\partial_{1\dot 1}$ (we ignore the zero-momentum constraint for now). 
This excitation has $K_3 = K_4 = K_5 = 1$, and the Bethe equations read
\begin{equation}
  1 = q^L \frac{u_3-\tilde{u}_4-i/2}{u_3-\tilde{u}_4+i/2} = q^L\frac{u_5-\tilde{u}_4-i/2}{u_5-\tilde{u}_4+i/2}\, ,
\end{equation}
and
\begin{equation} 
  \left( \frac{\tilde{u}_4+i/2}{\tilde{u}_4-i/2} \right)^L
    = q^{-2L} \frac{\tilde{u}_4-u_3-i/2}{\tilde{u}_4-u_3+i/2}\frac{\tilde{u}_4-u_5-i/2}{\tilde{u}_4-u_5+i/2}\, .
\end{equation}
At large $q$ we see that we must have $u_{3/5} - \tilde{u}_4 - i/2 \sim q^{-L}$. 
If we repeat this exercise for the remaining types of derivatives (i.e.~by introducing an $u_2$ 
and/or $u_6$ root) we find the same scaling for the $u_{3}$ and $u_5$ roots, while no scaling is
necessary for the $u_2$ and $u_6$ roots.

Let us assume that this is the general
pattern for states with an arbitrary number of excitations, and set
\begin{equation}
  u_{3,j} = \tilde{u}_{4,j} + i/2 + iq^{-L}\beta_{3,j}\, , \qquad 
  u_{5,j} = \tilde{u}_{4,j} + i/2 + iq^{-L}\beta_{5,j}\, ,
  \label{eq:fishnet-scaling}
\end{equation}  
for $j=1,\ldots, K_\partial$. It is now straightforward to take the
$q\to\infty$ limit of the Bethe equations. We find
\begin{equation}
  \alpha_{4,k}^L = (-1)^{K_R-1}
    \prod_{j=1}^{K_\partial}\frac{(\tilde{u}_{4,j}+3i/2)(\tilde{u}_{4,j}-i/2)}{(\tilde{u}_{4,j}+i/2)(\tilde{u}_{4,j}+i/2)}
\end{equation}
for the main roots, and
\begin{align}
  1 &= \prod_{j\neq k}^{K_2}\frac{u_{2,k}-u_{2,j}-i}{u_{2,k}-u_{2,j}+i}
        \prod_{j=1}^{K_\partial}\frac{u_{2,k}-\tilde{u}_{4,j}}{u_{2,k}-\tilde{u}_{4,j}-i}\, , \\
  1 &= \prod_{j\neq k}^{K_6}\frac{u_{6,k}-u_{6,j}-i}{u_{6,k}-u_{6,j}+i}
        \prod_{j=1}^{K_\partial}\frac{u_{6,k}-\tilde{u}_{4,j}}{u_{6,k}-\tilde{u}_{4,j}-i}\, , \\
  \left(\frac{\tilde{u}_{4,k}+i/2}{\tilde{u}_{4,k}-i/2}\right)^{L-K_R}
    &= \left(\frac{\tilde{u}_{4,k}+i/2}{\tilde{u}_{4,k}+3i/2}\right)^{K_R}
       \prod_{j\neq k}^{K_\partial}\frac{\tilde{u}_{4,k}-\tilde{u}_{4,j}-i}{\tilde{u}_{4,k}-\tilde{u}_{4,j}+i}\nonumber\, ,\\
    &\qquad \times\prod_{j=1}^{K_2}\frac{\tilde{u}_{4,k}-u_{2,j}+i}{\tilde{u}_{4,k}-u_{2,j}}
       \prod_{j=1}^{K_6}\frac{\tilde{u}_{4,k}-u_{6,j}+i}{\tilde{u}_{4,k}-u_{6,j}}
\end{align}
for the auxiliary roots. Since $E$ is independent of the $\tilde{u}_{4,j}$, it is natural
to include them among the auxiliary roots. Here we have used the equations at nodes
3 and 5 to eliminate the $\beta_{3,j}$ and $\beta_{5,j}$, which has the effect of changing the $\tilde{u}_{4}$
self-scattering term from the $\mathfrak{su}(2)$ form to the $\mathfrak{sl}(2)$ form appropriate for derivatives.

In order to check that we have scaled the auxiliary roots correctly we can compare our
Bethe equations to those derived in section \ref{sec:fishnetderivative}.
If we set $K_2 = K_6 = 0$ and identify
\begin{equation}
  e^{ip_{\partial,j}} = \frac{\tilde{u}_{4,j}+i/2}{\tilde{u}_{4,j}-i/2}\, ,\qquad
  e^{ip_{\phi,j}} = \alpha_j\, ,
\end{equation}
we see that the equations are identical.

Equipped with a guideline on how to scale the roots, we are now able to study two interesting sectors of the strongly $\beta$-twisted model \eqref{eq:Lagrangianbeta}.

\section{\mathversion{bold}Strongly \texorpdfstring{$\beta$}{}-Twisted Theory}

\subsection{\mathversion{bold}A Broken \texorpdfstring{$\mathfrak{su}(3)$}{} Sector}
\label{sec:su3}

The second simplest sector of strongly $\beta$-twisted theory, after the broken $\mathfrak{su}(2)$ sector described in section \ref{subsec:xy}, consists of the three scalars $\{\phi_1,\phi_2,\phi_3^\dagger\}$. In memory of the original $\mathcal{N}=4$ model we will call it ``broken $\mathfrak{su}(3)$ sector''.
One of these scalars has to be a conjugate scalar in order to avoid eclectic field content, which would result in a nilpotent Hamiltonian as described in section \ref{sec:eclectic}.
We choose $\phi_1$ as the vacuum and $\phi_2$ and $\phi_3^\dagger$ as (right-moving) excitations.
We start with the Beisert-Roiban equations in the so-called ``Beauty'' grading given in appendix 
\ref{app:beauty-bethe-equations} 
with a total of $K_4$ excitations $\{\phi_2,\phi_3^\dagger\}$, and $K_3$ excitations of $\phi_3^\dagger$ type.
The scaling of the momentum-carrying roots was discussed in the last chapter and according to \eqref{eq:ufour-scaling} for right-movers we have
\begin{equation}
u_{4,j}=-i/2 - iq^{-2}\alpha_{4,j}\, ,
\label{eq:ufourscalingsu3}
\end{equation}
where $u_4$ are the usual momentum-carrying Bethe roots and $\alpha=e^{ip}$ is our parametrization of the lattice momentum.
To cancel the remaining factors of $q$ in the twisted Beisert-Roiban equations we see that one option is to let
\begin{equation}
u_{3,j}=-iq^{-2}\alpha_{3,j}\, .
\label{eq:uthreescalingsu3}
\end{equation}
We then obtain the following Bethe equations
\begin{align}
\label{eqsu3,1}
\alpha_{4,k}^L&=(-1)^{K_4-1} \prod_{j=1}^{K_3} \frac{1}{\alpha_{4,k}-\alpha_{3,j}}\, ,\\
(-1)^{K_3-1}&=\prod_{j=1}^{K_4}(\alpha_{4,j}-\alpha_{3,k}),\\
1&=\prod_{j=1}^{K_4}\alpha_{4,j}\, ,\\
E&=-2\sum_{j=1}^{K_4}\frac{1}{\alpha_{4,j}}\, .
\label{eqsu3,2}
\end{align}
These equations are a generalization of the ones for the broken $\mathfrak{su}(2)$ sector \eqref{eq:cXYdispersion} - \eqref{eq:cXYquantization}.
In contrast, however, they are no longer obviously solvable. At the same time, they nevertheless look simpler than ``usual'' one-loop Bethe equations. It would be interesting to find an analytic solution procedure. To check these equations, we can use the twisted R-matrix from \cite{Beisert:2005if} to apply a nested algebraic Bethe ansatz as described in \cite{Essler:1992he}.
The scaled version of this R-matrix is
\begin{equation}
\label{eq:rmatrix}
R(u)=u\, P^-+i\, \mathcal{P}\, ,
\end{equation}
where $P^-$ is the projection operator on letters in chiral order, and $\mathcal{P}$ is the permutation operator.
One checks that this is the correct R-matrix by computing the Hamiltonian in the usual fashion \cite{Faddeev:1996iy}
\begin{equation}
H=-2i\frac{d\log T}{du}\bigg\rvert_{u=0},
\end{equation}
which leads to agreement with our Hamiltonian density given in equation \eqref{eq:Dbetascaled}.
We used this R-matrix and the corresponding transfer matrix to apply an algebraic Bethe ansatz of the form described in \cite{Essler:1992he}.
As a successful consistency check, we managed to rederive the Bethe equations \eqref{eqsu3,1} - \eqref{eqsu3,2}.

Since the sector consisting of \{$\phi_1,\phi_2,\phi_3$\} is eclectic, we know that it has a nilpotent Hamiltonian, see section \ref{sec:eclectic}.
Thus, one might suspect the Bethe ansatz to fail. However, it is interesting to observe the problems that arise.
The construction of an R-matrix and a monodromy works.
However, in the algebraic Bethe ansatz one uses so called RTT relations to determine a set of fundamental commutation relations between the matrix elements of the monodromy.
In this sector, the R-matrix has some zero entries, such that the RTT relations do not produce a complete set of these fundamental commutation relations. Therefore, a Bethe ansatz, at least in its standard form, is not consistent.

As mentioned above, we did not manage to explicitly solve \eqref{eqsu3,1} - \eqref{eqsu3,2} for a general number of excitations.
However, for one excitation of each type $\phi_2$ and $\phi_3^\dagger$ at any length $L$, the equations are simple enough to be solved exactly. We find that one of the momentum-carrying Bethe roots is from either of the two sets
\begin{align}
u_1\in &\{\exp((2k+1)i\pi/(L+1))\vert 0 \leq k \leq (L-1)/2\}\,\, \text{or} \\
u_1\in &\{\exp(2ki\pi/(L-1))\vert 1 \leq k \leq (L-2)/2 \} .
\end{align}
The other momentum-carrying Bethe root is its complex conjugate, and hence their sum, the energy, is real.
In general these two sets are disjoint, however an interesting phenomenon emerges when $L=4n+1$ for some integer $n$.
In this case the Bethe root $i$ appears in both sets, and thus the energy of the corresponding state is  $E=-2(-i+i)=0$.
Through explicit calculations we find that in this case a Jordan block of size two forms in the spectrum, with generalized eigenvalue zero. One of the two states in the block is a true eigenstate, and does correspond to this particular solution of the Bethe equations. Thus, in contrast to the derivative case discussed in \ref{sec:completeness}, here the proper eigenstate of the Jordan block is found by the Bethe ansatz. However, the remaining part of the Jordan block is still undetermined by this ansatz.

\subsection{\mathversion{bold}A Broken \texorpdfstring{$\mathfrak{su}(2\vert 3)$}{} Sector}
\label{sec:su23}
Within the $\beta$-twisted model, we can extend the sector from the last section by fermions $\{\bar\psi_{3,\dot{1}} ,\bar\psi_{3,\dot{2}} \}$ to what we call, once more keeping the connection with unscaled $\mathcal{N}$=4 SYM, a broken $\mathfrak{su}(2\vert 3)$ sector. Take again $\phi_1$ as the vacuum, and consider the excitations \{$\phi_2,\phi_3^\dagger ,\bar\psi_{3,\dot{1}} ,\bar\psi_{3,\dot{2}} \}$.
According to appendix \ref{app:twisted-bethe-equations} this corresponds to exciting the roots from $K_1$ through $K_4$ in the ``Beauty'' grading.
This is the sector that we claim to be equivalent to the one considered in \cite{Caetano:2016ydc}.
However, in \cite{Caetano:2016ydc} the so-called ABA grading is used, and the number of fermionic excitations is fixed explicitly to zero.
This implies that the sector considered in \cite{Caetano:2016ydc} reduces to a sector that is equivalent to the broken $\mathfrak{su}(3)$ case discussed in the last section, even though this is not immediately manifest in the form of the equations.
Indeed, as mentioned earlier, we were able to match the one-loop limit of their equations with \eqref{eqsu3,1}-\eqref{eqsu3,2}

The scaling of the $u_4$ and $u_3$ roots is identical to the one in the broken $\mathfrak{su}(3)$ case, and all other roots do not need to be scaled.
Plugging these roots into the Bethe equations in the Beauty grading given in appendix \ref{app:twisted-bethe-equations} we find
\begin{align}
1&=\prod_{j=1}^{K_4} \alpha_{4,j} \label{eq:BEsu23first}\, ,\\
1&=\prod_{j\neq k}^{K_1}\frac{u_{1,k}-u_{1,j}-i}{u_{1,k}-u_{1,j}+i} \prod_{j=1}^{K_2}\frac{u_{1,k}-u_{2,j}+i/2}{u_{1,k}-u_{2,j}-i/2}\label{eq:BEsu23second}\, ,\\
\left(\frac{u_{2,k}+i/2}{u_{2,k}-i/2} \right)^{K_3}&=\prod_{j=1}^{K_1}\frac{u_{2,k}-u_{1,j}+i/2}{u_{2,k}-u_{1,j}-i/2}\label{eq:BEsu23third}\, ,\\
1&=\left(-1\right)^{K_3-1} \prod_{j=1}^{K_4}\left(\alpha_{4,k}-\alpha_{3,j} \right) \prod_{j=1}^{K_2}\frac{u_{2,j}+i/2}{u_{2,j}-i/2}\label{eq:BEsu23fourth}\, ,\\
\alpha_{4,k}^L&=\left(-1\right)^{K_4-1} \prod_{j=1}^{K_3} \frac{1}{\alpha_{4,k}-\alpha_{3,j}}\, ,\label{eq:BEsu23fifth}\\
E&=-\sum_{j=1}^{K_4}\frac{2}{\alpha_{4,j}}\, .\label{eq:BEsu23sixth}
\end{align}
We observe a curious decoupling of these equations, in the sense that \eqref{eq:BEsu23second} and \eqref{eq:BEsu23third} may be solved independently of the remaining ones, and then be used as ``source terms'' for the latter.
The remaining Bethe equations look strikingly similar to \eqref{eqsu3,1} and \eqref{eqsu3,2} from the last section.
In fact, they are identical except for the additional last factor in \eqref{eq:BEsu23fourth}.
The reason for this phenomenon becomes clearer when looking at the action of $H$.
In fact, in sectors without derivatives, like the one we are considering here, $H$ does not distinguish between $\phi_3^\dagger$ and $\bar\psi_{3,\dot{\alpha}}$, as can be determined from the Hamiltonian density \eqref{eq:Dbetascaled}.
However interchanging positions of a $\phi_3^\dagger$ and a $\bar\psi_{3,\dot{\alpha}}$ yields two distinguishable states.

As an illustrative example let us compare the spin chain given by only one state $\tr(\phi_1\phi_3^\dagger \phi_3^\dagger)$ and the spin chain given by the two states $\ket{1}=\tr(\phi_1\phi_3^\dagger \bar\psi_{3,\dot{1}})$ and $\ket{2}=\tr(\phi_1\bar\psi_{3,\dot{1}}\phi_3^\dagger )$.
For the second spin chain we have two eigenstates $\ket{1}+\ket{2}$ and $\ket{1}-\ket{2}$ with eigenvalues $-2$ and $+2$ respectively.
Since for our first spin chain interchanging the two excitations gives back the same state, a state equivalent to $\ket{1}-\ket{2}$ would be identically 0.
Hence, for this chain we have only one state, which is automatically an eigenstate with eigenvalue $-2$.
We conclude that the similarity between the broken $\mathfrak{su}(3)$ sector and the broken $\mathfrak{su}(2\vert3)$ sector is expected, due to the identical form of $H$, while the differences are due to the distinguishable nature of the excitations $\phi_3^\dagger$ and $\bar\psi_{3,\dot{\alpha}}$.

\section{\mathversion{bold}Remarks on Higher Loop Corrections}
In this paper, we have mostly focused on the one-loop structure of the spectrum. Let us
make a few preliminary remarks on the extension to higher loops.
The analysis we present here will also highlight a subtlety of the relationship between the Bethe equations derived in section 4.1 of \cite{Caetano:2016ydc} and our equations.
Twisted asymptotic all-loop Bethe equation were already given in \cite{Beisert:2005if}, so 
one should be able to follow the usual procedure of solving these equation perturbatively around
a given one-loop solution. Once wrapping and pre-wrapping\cite{Fokken:2013mza} sets
in, the analysis becomes, of course, more complicated. Here we will restrict to determining
the  scaling of the momentum-carrying roots  following the logic of section \ref{sec:scaling}. 

\subsection{\mathversion{bold}Scaling of Momentum-Carrying Roots}
\label{sec:rootscaling}

Let us thus consider a single right-moving excitation with momentum-carrying root
$u = u_{4,j}$. At generic twist $q$ the spin chain momentum is given by \cite{Beisert:2005if}
\begin{equation}
  e^{ip} := q^2 \frac{x^+}{x^-}\, ,
  \label{eq:p-all-loop}
\end{equation}
where $x^{\pm}$ are the weak coupling solutions to
\begin{equation}
  x^\pm + \frac{1}{x^\pm} = \frac{u\pm i/2}{g} \, , 
  \label{eq:Zhukovsky-def}
\end{equation}
i.e.
\begin{equation}
  x^\pm = \frac{u\pm i/2}{g}- \frac{g}{u\pm i/2} - \frac{g^3}{(u\pm i/2)^3} + O(g^5) \, .
\end{equation}
Our task is to find the appropriate scaling form of $u$, such that the coefficients of
$\xi^2$ in the weak coupling expansion of $e^{ip}$ have a finite $q\to\infty$ limit.
As in the one-loop case, we will parametrize $u$ by $\alpha$, but now $\alpha$ is a 
series in $\xi^2$:
\begin{equation}
  \alpha = \sum_{n=0}^\infty \alpha_{(n)} \xi^{2n}\, .
\end{equation}
We claim that the appropriate scaling\footnote{The scaling form of $u$ is unique up to a
  redefinition of $\alpha$.} is
\begin{align}
  u &= -i/2 -i \alpha q^{-2} + i\xi^2\alpha^{-1}\, , \\
   &= -i/2-i\alpha_{(0)} q^{-2} + \left( \frac{i}{\alpha_{(0)}} -i\alpha_{(1)} q^{-2}\right)\xi^2
    +\left(-\frac{i\alpha_{(1)}}{\alpha_{(0)}^2} - i\alpha_{(2)} q^{-2}\right)\xi^4
    + O(\xi^6)\, .
  \label{eq:u-all-loop}
\end{align}
Note that the leading term agrees with \eqref{eq:ufour-scaling}.
The solution is constructed exactly such that $x^+$ behaves nicely for large $q$, in fact
we have
\begin{equation}
  x^+ = -\frac{i}{\xi q}\alpha\, ,
\end{equation}
and it is then easy to see that $e^{ip}$ has a well-defined strong twisting limit,
order by order in $\xi^2$:
\begin{equation}
  q^2 \frac{x^+}{x^-} = \frac{\alpha}{1-\xi^2\alpha^{-1}} + O(q^{-2}).
  \label{eq:p-all-loop-from-alpha}
\end{equation}

It is paramount that the eigenvalue of $\delta\mathfrak{D}$, i.e. the anomalous dimension $\gamma$, also has a finite limit.
This is guaranteed, if all the contributions $\gamma_s$ to the anomalous dimension of the individual Bethe roots are finite and indeed
\begin{equation}
  \gamma_s := 2ig\left(\frac{1}{x^+}-\frac{1}{x^-}\right)
    = -2\xi^2\alpha^{-1} + O(q^{-2}) .
  \label{eq:gamma-all-loop}
\end{equation}
We note that \eqref{eq:p-all-loop}, \eqref{eq:p-all-loop-from-alpha} and \eqref{eq:gamma-all-loop}
together imply the following chiral dispersion law, to be compared with equation~(4.6) of \cite{Caetano:2016ydc}
\begin{equation}
  \gamma_s = \sqrt{1-4e^{-ip}\xi^2}-1\, .
  \label{eq:dispersion-all-loop}
\end{equation}

\subsection{\mathversion{bold}Comparison to Previous Work}
We take our underlying field theories to be defined by the strong twisting limit
of the perturbation series of twisted SYM. We thus expand in $g$ (equivalently $\xi$) first,
and then send $q\to\infty$.
In contrast, $\xi$ is kept finite as $q\to\infty$  in section 4.1 of \cite{Caetano:2016ydc},
and only after the strong twisting limit an expansion in $\xi^2$ is performed.
The order of limits is thus reversed in comparison to our approach.
It turns out that exchanging the order of limits leads to subtle differences.

To explicate this, let us redo the scaling analysis, but now
keeping $\xi$ finite as in \cite{Caetano:2016ydc}. 
For clarity we use hatted variables when taking the limits in `reverse' order.
The finiteness of \eqref{eq:p-all-loop} together with 
(this equation follows immediately from \eqref{eq:Zhukovsky-def})
\begin{equation}
  \hat{x}^+ + \frac{1}{\hat{x}^+} - \hat{x}^- - \frac{1}{\hat{x}^-} = \frac{i}{g}
\end{equation}
imply the scaling
\begin{equation}
  \hat{x}^+ \sim q^{-1}, \qquad \hat{x}^- \sim q \, .
\end{equation}
From \eqref{eq:Zhukovsky-def} we then immediately find (compare with Eq.~(4.9)
 of Ref.~\cite{Caetano:2016ydc})
\begin{equation}
  \hat{x}^+ = \frac{\xi}{q}\left(\frac{1}{\hat{u} + i/2}+O(q^{-2})\right), \qquad 
  \hat{x}^- = \frac{q}{\xi}\bigr(\hat{u}-i/2 + O(q^{-2})\bigl) ,
\end{equation}
and the single-magnon anomalous dimension is
\begin{equation}
  \hat{\gamma}_s := 2ig\left(\frac{1}{\hat{x}^+}-\frac{1}{\hat{x}^-}\right)
   = 2i(\hat{u}+i/2) + O(q^{-2}) .
\end{equation}
We can compare with the usual perturbative order by identifying $\hat{\gamma}_s = \gamma_s$,
which leads to
\begin{equation}
  \hat{u} = -i/2 + i\xi^2\alpha^{-1}\, .
\end{equation}
Referring back to \eqref{eq:u-all-loop}, we see that the relationship between $u$ and $\hat u$
is  non-trivial. Let us finally mention that the Bethe equation derived
in \cite{Caetano:2016ydc} agrees with (a subsector of) those of Sec.~\ref{sec:su23}, once this
relationship is taken into account. It is thus possible that exchanging the order of limits 
only leads to `superficial' differences, resulting in identical results for physical
quantities.

\section{\mathversion{bold}Conclusions and Outlook}
\label{sec:conclusions}
We have seen that the strongly twisted models appear to be, in comparison with the original $\mathcal{N}$=4 model, both much simpler in some ways as well as much more complicated in others. The former, because the number of Feynman diagrams governing their perturbative expansion is vastly reduced \cite{Gurdogan:2015csr,Sieg:2016vap,Caetano:2016ydc,Chicherin:2017cns,Gromov:2017cja,Chicherin:2017frs,Grabner:2017pgm,Kazakov:2018hrh, Gromov:2018hut}. The latter, since their dilatation operator ceases to be diagonalizable. It therefore seems to us that the twisted models are, all in all, neither simpler nor more complicated, but simply rather {\it different} from their mother theory.

The main reason for the different nature of the twisted models is their non-unitarity. An immediate consequence is the non-hermiticity of the dilation generator, which we have worked out explicitly at one-loop-order for the strongly twisted models, cf.~\eqref{eq:Dbetascaled}, \eqref{eq:Dfishnet}, \eqref{eq:Dgammascaled}. Already at this leading order it may no longer be fully diagonalized. From basic linear algebra, the best thing one can do is to bring the non-diagonalizable sectors in Jordan normal form. This was in some special cases already noticed and briefly discussed in \cite{Caetano:2016ydc}, and in more detail (but still for the special case) in \cite{Gromov:2017cja}, where a connection to logarithmic conformal field theory was made. In this paper we have shown many more examples for non-diagonalizable states of these models. Perhaps surprisingly, even in the (chiral) $\phi_1, \phi_2$-sector of the fishnet model {\it with} derivatives a Jordan block structure appears, which seems to be a novel result, cf.~section \ref{sec:completeness}. For the states of this sector not part of a Jordan block we proposed novel Bethe equations, see sections \ref{sec:fishnetderivative} and \ref{sec:fishnet-scaling}.

The lack of complete diagonalizability of the strongly twisted models should not be taken lightly. After all, one way to state the meaning of quantum integrability is the simultaneous diagonalizability of an infinite set of charges in involution. What if none of them may be diagonalized in the first place? This certainly obscures the very meaning of ``integrability'' on a theoretical level. Recall that one of the motivations to study the twisted models has been to get a useful insight into the reasons underlying the integrability of certain planar four-dimensional quantum field theories, with the hope of subsequently transporting these insights to full-fledged $\mathcal{N}$=4 SYM.

The main conceptual purpose of this paper has then been to demonstrate the significant differences between the $\mathcal{N}$=4 theory and the twisted models that appear, in the example of the spectral problem, already at the leading one-loop level.  Let us remember that in the case of $\mathcal{N}$=4 SYM the careful analysis of the precise one-loop structure serves as the solid basis for the higher-loop asymptotic Bethe ansatz (ABA) and provides the necessary ``initial conditions'' for the exact functional equations of the quantum spectral curve (QSC) (for recent reviews of the latter, including a discussion of the twisted models, see \cite{Kazakov:2018hrh,Gromov:2017blm}). It is surely fair to ask in what sense the QSC provides an ``exact solution'' of the spectral problem of a model, if a large fraction of its operators cannot be diagonalized in the first place.

On the bright side, the novel integrable models of \cite{Gurdogan:2015csr} pose an interesting mathematical challenge for the future: How can one systematically and fully adapt the quantum inverse scattering method to integrable models with a non-hermitian (and non-pseudo-hermitian) Hamiltonian? How can one completely bring this Hamiltonian in Jordan normal form, generalizing or suitably replacing the Bethe ansatz? Are there Jordan cells with generalized eigenvalues different from zero? If so, how to find these generalized eigenvalues? What happens to the Jordan cells once one takes higher loop corrections into account? Note that the example of a cell studied in chapter 7.2 of \cite{Gromov:2017cja} was argued to stay intact at every order in perturbation theory. We also feel that the interesting connections to logarithmic conformal field theory pointed out in \cite{Gromov:2017cja} should be systematically explored. Finally, it would be very interesting to find a physical application for the strongly twisted models. 

\section*{Acknowledgments}
\label{sec:acknowledgements}
We are thankful to João Caetano, Vladimir Kazakov, Gregory Korchemsky, Florian Loebbert, Dennis Müller and Stijn van Tongeren for inspiring discussions. We thank J.~Caetano, V.~Kazakov and G.~Korchemsky for excellent talks on the subject. The work of ACI is supported by the Villum Foundation. LZ has been supported by the DFG-funded graduate school {\it GK 1504 Masse, Spektrum, Symmetrie} in the preparatory phase of this project.

\appendix

\section{\mathversion{bold}Dilatation Operator of 
Strongly \texorpdfstring{$\gamma_i$}{}-Twisted Models}
\label{app:Dilatation_gamma}

In section \ref{subsec:dilatation} we have derived the one-loop dilatation operator of the two strongly twisted theories that we investigate in this paper, namely the strongly $\beta$-twisted model and the fishnet model.
The same arguments that led to the one-loop dilatation operator in these theories can also be applied to the more general deformations of strongly $\gamma_i$-twisted models where all three double-scaled couplings $\xi_i$ are a priori distinct and not necessarily zero. Let us parameterize the different coupling constants $\xi_i$ by setting $\xi_i=\xi a_i$ for some reference coupling $\xi$.
As before we divide the quantum corrections to the dilatation operator as $\delta\mathfrak{D}=\xi^2H+\mathcal{O}(\xi^4)$. The one-loop part acquires additional factors compared to the $\beta$-twisted version given in \eqref{eq:Dbetascaled}, but stays structurally the same.
It is given by
\begin{equation}
\label{eq:Dgammascaled}
(\mathcal{H}_{n,n+1}^{\rm{s} \gamma \rm{t}})_{\lett_n \lett_{n+1}}^{\lett'_n \lett'_{n+1}}=c(a_1,a_2,a_3)\,(\mathcal{H}_{n,n+1}^{\rm{s} \beta \rm{t}})_{\lett_n \lett_{n+1}}^{\lett'_n \lett'_{n+1}}\, ,
\end{equation}
where $c(a_1,a_2,a_3)$ depends on the fermionic or bosonic nature of the exchanged flavors.
We give it case by case.
\begin{itemize}
\item Case 1: For an exchange of two scalars $c=a_i^2$, where the subscript $i$ corresponds to the flavor {\it not} taking part in the exchange.
For example, if the dilatation operator exchanges $\phi_1$ and $\phi_2$, then $c=a_3^2$.
\item Case 2: For an exchange of two fermions $c=a_i a_j$, where the subscripts $i,j$ correspond to the flavors that are exchanged.
\item Case 3: For an exchange of a fermion and a scalar $c=a_i a_j$, where the subscripts correspond to the flavor of the fermion and the flavor not taking part in the exchange.
\item Case 4: If the fermionic or bosonic nature of the excitations changes during the flavor exchange, $c$ is the square root of the product of two of the $c$'s from the previous cases.
The factors are taken such that one factor $\sqrt{c}$ corresponds to the initial configuration and the other factor of $\sqrt{c}$ corresponds to the configuration after the flavor exchange.
To illustrate this let us consider the matrix element of the dilatation operator corresponding to $\partial \phi_1,\phi_2^\dagger \rightarrow \bar{\psi}_2, \psi_1$, where we suppressed spinor indices. The $c$ in \eqref{eq:Dgammascaled} for this example is the product of $\sqrt{c}=a_3$ from case 1 and $\sqrt{c}=\sqrt{a_1 a_2}$ from case 2.
\end{itemize}

The above analysis works for those special cases where some of the $a_i$ are zero.
This includes the fishnet model for which $a_1=a_2=0$ and $a_3=1$.
However, one can construct even more general double scaled theories by taking the limit $q_i\rightarrow 0$, with $g/q_i$ kept fixed for some of the twist angles.
In this case, the dilatation operator is again distinct from the one described in this section, and one has to go through similar calculations to obtain it.

\section{\mathversion{bold}Twisted One-Loop Bethe Equations}
\label{app:twisted-bethe-equations}
In this appendix, we write down the twisted Bethe equations from \cite{Beisert:2005if} that
we need. The Dynkin diagram of $\mathfrak{su}(2,2|4)$ admits various gradings, which leads
to different sets of Bethe equations. Here we will use the ``Beauty'' grading\cite{Beisert:2003yb}
and the ABA grading\cite{Beisert:2005fw}.

\subsection{\mathversion{bold}``Beauty'' Grading}
\label{app:beauty-bethe-equations}
\begin{table}[h]
\begin{center}
{\renewcommand{\arraystretch}{1.2} 
\begin{tabular}{c|cc|cc}
    &4                       &5                       &6                     &7                     \\ \hline
  4 &$\phi_2$                &$\phi_3$                &$\psi_{41}$           &$\psi_{42}$           \\
  3 &$\phi_3^\dagger$        &$\phi_2^\dagger$        &$\psi_{11}$           &$\psi_{12}$           \\ \hline
  2 &$\bar{\psi}_{3\dot{1}}$ &$\bar{\psi}_{2\dot{1}}$ &$\partial_{1\dot{1}}$ &$\partial_{2\dot{1}}$ \\
  1 &$\bar{\psi}_{3\dot{2}}$ &$\bar{\psi}_{2\dot{2}}$ &$\partial_{1\dot{2}}$ &$\partial_{2\dot{2}}$ \\
\end{tabular}}
\end{center}
\caption{Single excitations of the full $\mathcal{N} = 4$ SYM spin chain in the ``Beauty'' grading. The table should
be read as follows: Consider an state with the non-zero $K$s being $K_{j}=K_{j+1}= \cdots = K_{k} = 1$,
where $1\leq j \leq 4 \leq k \leq 7$. The corresponding excitation, over a vacuum of $\phi_1$s, is the one
listed at row $j$ and column $k$.}
\label{tab:beauty-excitations}
\end{table}
For this grading we only consider the $\beta$-twist. We
thus set $\gamma_1=\gamma_2=\gamma_3=\beta$ and $q = e^{-i\beta/2}$.
The twisted Bethe equations are\cite{Beisert:2005if}:

\begin{align}
  1 &= q^{2K_4-4K_5+2K_6} \prod_{j=1}^{K_4}\frac{u_{4,j}+i/2}{u_{4,j}-i/2}\, , \label{eq:beautymomconstr}\\
  1 &= \prod_{j\neq k}^{K_1}\frac{u_{1,k}-u_{1,j}-i}{u_{1,k}-u_{1,j}+i}
       \prod_{j=1}^{K_2}\frac{u_{1,k}-u_{2,j}+i/2}{u_{1,k}-u_{2,j}-i/2}\, , \\
  1 &= \prod_{j=1}^{K_1}\frac{u_{2,k}-u_{1,j}+i/2}{u_{2,k}-u_{1,j}-i/2}
       \prod_{j=1}^{K_3}\frac{u_{2,k}-u_{3,j}-i/2}{u_{2,k}-u_{3,j}+i/2}\, , \\
  1 &= q^{2K_4-4K_5+2K_6}\prod_{j=1}^{K_2}\frac{u_{3,k}-u_{2,j}-i/2}{u_{3,k}-u_{2,j}+i/2}\nonumber \\
    &\qquad\times\prod_{j\neq k}^{K_3}\frac{u_{3,k}-u_{3,j}+i}{u_{3,k}-u_{3,j}-i} 
    \prod_{j=1}^{K_4}\frac{u_{3,k}-u_{4,j}-i/2}{u_{3,k}-u_{4,j}+i/2}\, , \\
  \left(\frac{u_{4,k}+i/2}{u_{4,k}-i/2}\right)^L &= q^{-2L-2K_3+6K_5-4K_6} 
       \prod_{j=1}^{K_3}\frac{u_{4,k}-u_{3,j}-i/2}{u_{4,k}-u_{3,j}+i/2} \nonumber \\
    &\qquad \times\prod_{j\neq k}^{K_4} \frac{u_{4,k}-u_{4,j}+i}{u_{4,k}-u_{4,j}-i}
    \prod_{j=1}^{K_5}\frac{u_{4,k}-u_{5,j}-i/2}{u_{4,k}-u_{5,j}+i/2}\, , \\
  1 &= q^{4L+4K_3-6K_4+2K_6}\prod_{j=1}^{K_4}\frac{u_{5,k}-u_{4,j}-i/2}{u_{5,k}-u_{4,j}+i/2} \nonumber \\
    &\qquad \times\prod_{j\neq k}^{K_5}\frac{u_{5,k}-u_{5,j}+i}{u_{5,k}-u_{5,j}-i}
    \prod_{j=1}^{K_6}\frac{u_{5,k}-u_{6,j}-i/2}{u_{5,k}-u_{6,j}+i/2}\, ,\\
  1 &= q^{-2L-2K_3+4K_4-2K_5}\prod_{j=1}^{K_5}\frac{u_{6,k}-u_{5,j}-i/2}{u_{6,k}-u_{5,j}+i/2} \nonumber \\
    &\qquad\times\prod_{j=1}^{K_7}\frac{u_{6,k}-u_{7,j}+i/2}{u_{6,k}-u_{7,j}-i/2}\, , \\
  1 &= \prod_{j=1}^{K_6}\frac{u_{7,k}-u_{6,j}+i/2}{u_{7,k}-u_{6,j}-i/2}
    \prod_{j\neq k}^{K_7}\frac{u_{7,k}-u_{7,j}-i}{u_{7,k}-u_{7,j}+i}\, .
\end{align}
The elementary excitations are listed in Table \ref{tab:beauty-excitations}.
The momentum constraint \eqref{eq:beautymomconstr} agrees with Eq.~\eqref{eq:genmomconstr},
where $K_R = K_4-K_5$ and $K_L = K_5-K_6$.
\newpage

\subsection{\mathversion{bold}ABA Grading}
\label{app:ABA-bethe-equations}
\begin{table}[h]
\begin{center}
{\renewcommand{\arraystretch}{1.2} 
\begin{tabular}{c|c|cc|c}
    &4                       &5                       &6                     &7                       \\ \hline
  4 &$\phi_2$                &$\psi_{41}$             &$\psi_{42}$           &$\phi_3$                \\ \hline
  3 &$\bar{\psi}_{3\dot{1}}$ &$\partial_{1\dot{1}}$ &$\partial_{2\dot{1}}$   &$\bar{\psi}_{2\dot{1}}$ \\ 
  2 &$\bar{\psi}_{3\dot{2}}$ &$\partial_{1\dot{2}}$ &$\partial_{2\dot{2}}$   &$\bar{\psi}_{2\dot{2}}$ \\ \hline
  1 &$\phi_3^\dagger$        &$\psi_{11}$           &$\psi_{12}$             &$\phi_2^\dagger$        \\
\end{tabular}} \hspace{.5cm}
{\renewcommand{\arraystretch}{1.2} 
\begin{tabular}{c|c|cc|c}
    &4                       &5                       &6                     &7                       \\ \hline
  4 &$\phi_3^\dagger$        &$\psi_{11}$           &$\psi_{12}$             &$\phi_2^\dagger$        \\ \hline
  3 &$\bar{\psi}_{3\dot{1}}$ &$\partial_{1\dot{1}}$ &$\partial_{2\dot{1}}$   &$\bar{\psi}_{2\dot{1}}$ \\ 
  2 &$\bar{\psi}_{3\dot{2}}$ &$\partial_{1\dot{2}}$ &$\partial_{2\dot{2}}$   &$\bar{\psi}_{2\dot{2}}$ \\ \hline
  1 &$\phi_2$                &$\psi_{41}$             &$\psi_{42}$           &$\phi_3$                \\ 
\end{tabular}}
\end{center}
\caption{Single excitations of the full $\mathcal{N} = 4$ SYM spin chain in the ABA grading. The
left table is in the conventions of \cite{Beisert:2005if}, while the right is the R-symmetry rotated
variant  we use in connection with the $\beta$-twist. The notation
is the same as Table \ref{tab:beauty-excitations}.}
\label{tab:aba-excitations}
\end{table}
\begin{table}[h]
\begin{center}
\begin{tabular}{c|c|c}
       & $\gamma_3$-twist & $\beta$-twist      \\ \hline
 $t_0$ & $-K_1-K_3 + 2K_4 - K_5 - K_7$ & $2K_4-2K_5-2K_7$ \\
 $t_1$ & $L+K_3-2K_4+K_5$ & $-2K_4+2K_5+2K_7$  \\
 $t_3$ & $L-K_1-K_7$      & $0$                \\
 $t_4$ & $-2L+2K_1+2K_7$  & $-2L+2K_1+2K_7$    \\
 $t_5$ & $L-K_1-K_7$      & $2L-2K_1-2K_7$     \\
 $t_7$ & $L+K_3-2K_4+K_5$ & $2L-2K_1-2K_4+2K_5$
\end{tabular}
\end{center}
\caption{Twist factors for the ABA Bethe equation.}
\label{eq:ABA-twist-factors}
\end{table}

For the Bethe equations in the ABA grading we will consider two different twists. For the $\gamma_3$-twist
we set $\gamma_1=\gamma_2=0$ and $q= e^{-i\gamma_3/2}$.
For the $\beta$-twist we set $\gamma_1=\gamma_2=\gamma_3=\beta$ and $q = e^{-i\beta/2}$.
The corresponding values of $t_0,\ldots,t_7$ are given in Table \ref{eq:ABA-twist-factors}.
The elementary excitations are listed in Table \ref{tab:aba-excitations}. Note that
for the $\beta$-twist we have performed an R-symmetry rotation compared to the 
conventions of \cite{Beisert:2005if}.

The zero-momentum constraint for the $\beta$-twist takes the form \eqref{eq:genmomconstr}
with $K_R = K_4 - K_5$ and $K_L = K_7$. For the $\gamma_3$ twist we focus on the 
elementary excitations $\{\phi_2,\phi_2^\dagger,\partial_{\alpha\dot\alpha}\}$, since
these are the one that do not decouple in the fishnet limit. This leads to 
the restrictions $K_3 = K_5$ and $K_1 = K_7$, cf.~Table \ref{tab:aba-excitations}.
We then again find the zero-momentum constraint to be of the form \eqref{eq:genmomconstr},
with $K_R = K_4 - K_3 = K_4 - K_5$ and $K_L = K_1 = K_7$.

In the ABA grading the Bethe equations take the following form\cite{Beisert:2005if}:
\begin{align}
  1 &= q^{t_0} \prod_{j=1}^{K_4}\frac{u_{4,j}+i/2}{u_{4,j}-i/2}\, , \\
  1 &=  q^{t_1}\prod_{j=1}^{K_2}\frac{u_{1,k}-u_{2,j}+i/2}{u_{1,k}-u_{2,j}-i/2}\, , \\
  1 &= \prod_{j=1}^{K_1}\frac{u_{2,k}-u_{1,j}+i/2}{u_{2,k}-u_{1,j}-i/2} \nonumber \\
    &\qquad\times   \prod_{j\neq k}^{K_2}\frac{u_{2,k}-u_{2,j}-i}{u_{2,k}-u_{2,j}+i}
       \prod_{j=1}^{K_3}\frac{u_{2,k}-u_{3,j}+i/2}{u_{2,k}-u_{3,j}-i/2}\, , \\
  1 &= q^{t_3}\prod_{j=1}^{K_2}\frac{u_{3,k}-u_{2,j}+i/2}{u_{3,k}-u_{2,j}-i/2}
       \prod_{j=1}^{K_4}\frac{u_{3,k}-u_{4,j}-i/2}{u_{3,k}-u_{4,j}+i/2}\, , \\
  \left(\frac{u_{4,k}+i/2}{u_{4,k}-i/2}\right)^L &= q^{t_4} 
       \prod_{j=1}^{K_3}\frac{u_{4,k}-u_{3,j}-i/2}{u_{4,k}-u_{3,j}+i/2} \nonumber \\
    &\qquad \times\prod_{j\neq k}^{K_4} \frac{u_{4,k}-u_{4,j}+i}{u_{4,k}-u_{4,j}-i}
    \prod_{j=1}^{K_5}\frac{u_{4,k}-u_{5,j}-i/2}{u_{4,k}-u_{5,j}+i/2}\, , \\
  1 &= q^{t_5}\prod_{j=1}^{K_4}\frac{u_{5,k}-u_{4,j}-i/2}{u_{5,k}-u_{4,j}+i/2} 
    \prod_{j=1}^{K_6}\frac{u_{5,k}-u_{6,j}+i/2}{u_{5,k}-u_{6,j}-i/2}\, ,\\
  1 &= \prod_{j=1}^{K_5}\frac{u_{6,k}-u_{5,j}+i/2}{u_{6,k}-u_{5,j}-i/2} \nonumber \\
    &\qquad\times\prod_{j\neq k}^{K_6}\frac{u_{6,k}-u_{6,j}-i}{u_{6,k}-u_{6,j}+i}
    \prod_{j=1}^{K_7}\frac{u_{6,k}-u_{7,j}+i/2}{u_{6,k}-u_{7,j}-i/2}\, , \\
  1 &= q^{t_7}\prod_{j=1}^{K_6}\frac{u_{7,k}-u_{6,j}+i/2}{u_{7,k}-u_{6,j}-i/2}\, .
\end{align}

The twisted Bethe equations in the ABA grading are also given in appendix C of \cite{Caetano:2016ydc}.
To match the conventions of \cite{Beisert:2005if}, as also employed in the present paper, 
we find it necessary to send $q_1 \to q_1^{-1}$ and $q_2 \to q_2^{-1}$ in the equations
of \cite{Caetano:2016ydc}.
For example, the zero-momentum constraint is given as
\begin{equation*}
  \prod_{k=1}^{K_4} \frac{x^+_{4,k}}{x^-_{4,k}} = q_2^{-2J_3}q_3^{-2J_2} ,
  \qquad\qquad\qquad \text{(Eq.~(C.1) of \cite{Caetano:2016ydc})}
\end{equation*}
with $x_{4,k}^\pm = g^{-1}(u_{4,k}\pm i/2) + O(g)$.
In the strongly $\beta$-twisted limit $q_1=q_2=q_3\to \infty$ this would imply that both 
$\phi_2$ and $\phi_3$ are right-movers. With our conventions $\phi_2$ is a right-mover,
but $\phi_3$ is a left-mover.

\section{\mathversion{bold}Derivatives in the Strongly \texorpdfstring{$\beta$}{beta}-Twisted Model}
\label{app:beta-with-derivatives}
In this section, we give Bethe equations for the sector of the strongly
$\beta$-twisted model consisting of the excitations 
$\{\phi_3^\dagger,\psi_{1\alpha},\bar{\psi}_{3\dot\alpha},\partial_{\alpha\dot\alpha}\}$.
This corresponds to the upper left part of Table \ref{tab:aba-excitations} (right).
We thus set $K_1=K_7=0$. The number of right-movers is $K_R = K_4 - K_5$, and
there are no left-movers. 

The derivation closely follows that of section \ref{sec:fishnet-scaling}, so we will
be brief. By considering single excitation states we conjecture the scaling
\begin{equation}
  u_{5,j} = i/2+\tilde{u}_{4,j}+iq^{-2L}\beta_j\, ,
\end{equation}
with all other auxiliary roots unscaled. Plugging this into the equations given
in appendix \ref{app:ABA-bethe-equations}, we find, in the $q\to\infty$ limit,
\begin{equation}
  \alpha_k^L = (-1)^{K_R-1}\prod_{j=1}^{K_3}\frac{u_{3,j}+i}{u_{3,j}}
               \prod_{j=1}^{K_5}\frac{\tilde{u}_{4,j}-i/2}{\tilde{u}_{4,j}+i/2}
\end{equation}
for the main roots, and
\begin{align}
  1 &= \prod_{j\neq k}^{K_2}\frac{u_{2,k}-u_{2,j}-i}{u_{2,k}-u_{2,j}+i}
       \prod_{j=1}^{K_3}\frac{u_{2,k}-u_{3,j}+i/2}{u_{2,k}-u_{3,j}-i/2}\, , \\
  1 &= \left(\frac{u_{3,k}}{u_{3,k}+i}\right)^{K_R}
       \prod_{j=1}^{K_2}\frac{u_{3,k}-u_{2,j}+i/2}{u_{3,k}-u_{2,j}-i/2}
       \prod_{j=1}^{K_5}\frac{u_{3,k}-\tilde{u}_{4,j}-i/2}{u_{3,k}-\tilde{u}_{4,j}+i/2}\, ,\\
  \left(\frac{\tilde{u}_{4,k}+i/2}{\tilde{u}_{4,k}-i/2}\right)^{L-K_R}
    &= \prod_{j=1}^{K_3}\frac{\tilde{u}_{4,k}-u_{3,j}-i/2}{\tilde{u}_{4,k}-u_{3,j}+i/2}
       \prod_{j=1}^{K_6}\frac{\tilde{u}_{4,k}-u_{6,j}+i}{\tilde{u}_{4,k}-u_{6,j}}\, , \\
  1 &= \prod_{j=1}^{K_5}\frac{u_{6,k}-\tilde{u}_{4,j}}{u_{6,k}-\tilde{u}_{4,j}-i}
       \prod_{j\neq k}^{K_6}\frac{u_{6,k}-u_{6,j}-i}{u_{6,k}-u_{6,j}+i}
\end{align}
for the auxiliary roots. As in section \ref{sec:fishnet-scaling} we were
able to eliminate the $\beta_{j}$ roots.

It is interesting to try to reproduce the above equations using the coordinate 
Bethe ansatz. We proceed along the same lines as in section \ref{sec:fishnetderivative}.
For simplicity we will restrict to states with only $u_4$ and $u_5$ roots.
This corresponds to only keeping the excitations $\phi_3^\dagger$ and $\psi_{11}$.
A simple calculation shows that the $S$-matrix is
\begin{align}
  S|\phi_3^\dagger(p_1)\phi_{3}^\dagger(p_2)\rangle
    &= -|\phi_3^\dagger(p_2)\phi_{3}^\dagger(p_1)\rangle, \\
  S|\phi_3^\dagger(p_1)\psi_{11}(p_2)\rangle
    &= e^{ip_2}|\psi_{11}(p_2)\phi_3^\dagger(p_1)\rangle, \\
  S|\psi_{11}(p_1)\psi_{11}(p_2)\rangle
    &= -S_{\psi,\psi}(p_1,p_2)|\psi_{11}(p_2)\psi_{11}(p_1)\rangle .\label{eq:Spsipsi}
\end{align}
The fermion-fermion $S$-matrix element cannot be fixed in the usual way
by imposing the eigenvalue equation on scattering states, even if one considers
more than two excitations. We thus leave it as an unspecified function.\footnote{As
  always for an $S$-matrix it should satisfy $S_{\psi,\psi}(p_1,p_2) = S_{\psi,\psi}(p_2,p_1)^{-1}$.}
It would be interesting to see whether higher loop corrections fix
$S_{\psi,\psi}(p_1,p_2)$.

Since the scattering is transmission diagonal, we can use \eqref{eq:BAfishnetder} to write
down the Bethe equations\footnote{The boundary conditions for the fermions
  introduce an additional sign factor in \eqref{eq:BAbetaferm} which cancels against
  the explicit sign in \eqref{eq:Spsipsi}},
\begin{align}
  e^{i p_{\phi,k} L} &= (-1)^{K_R-1}\prod_{j=1}^{K_5} e^{-ip_{\psi,j}}, \label{eq:BAbetascal}\\
  e^{i p_{\psi,k} L} &= e^{i p_{\psi,k} K_R} \prod_{j\neq k}^{K_5} S_{\psi,\psi}(p_{\psi,k},p_{\psi,j})^{-1} . \label{eq:BAbetaferm}
\end{align}
These agree with those we obtained above by scaling, if one sets $S_{\psi,\psi}(p_1,p_2) = 1$.
Note that this is also the value of the fermion-fermion $S$-matrix that follows from
twisting the Beisert $S$-matrix.\footnote{In fact, this particular matrix element of the
$S$-matrix is not affected by the $\beta$-twist.}

It might seem surprising that one is free to choose $S_{\psi,\psi}$. 
The explanation appears to be related to the fact 
that the system is very degenerate. Taking
the product of \eqref{eq:BAbetaferm} over all $k$ we get
\begin{equation}
  \left(\prod_{j=1}^{K_5}e^{-ip_{\psi,j}}\right)^{L-K_R} = 1\, .
  \label{eq:BAbetafermprod}
\end{equation}
Since the energy is expressed in terms of the $p_{\phi,k}$ only, it follows that
the only influence the fermion sector has on the spectrum is through
the choice of which root of unity from \eqref{eq:BAbetafermprod}
to insert into \eqref{eq:BAbetaferm} and the zero-momentum constraint.

\section{\mathversion{bold}Nilpotency Proof}
\label{app:nilpotency-proof}
In this appendix we prove the following \\
\textbf{Theorem:} Consider an operator $\tr(A_1\cdots A_L)$.
If there \emph{does not} exist a $b \in F$ such that (we use the notation introduced in section \ref{subsec:letters})
\begin{equation}
  a_i := \flav(A_i) \in \{b, b_+, \bar{b}_-\}, \qquad \text{for all $i = 1, \ldots, L$}\, ,
\end{equation}
then $H^N |A_1 \cdots A_L\rangle = 0$ for some $N > 0$. 
Here $H$ is the one-loop Hamiltonian of either the strongly $\beta$-twisted or fishnet model,
\begin{equation}
  H = \sum_{n=1}^L \mathcal{H}_{n,n+1}^{\text{s$\beta$t}}\text{ , or }
  H = \sum_{n=1}^L \mathcal{H}_{n,n+1}^{\text{FN}}\, .
\end{equation}
The Hamiltonian densities $H_{n,n+1}^{\text{s$\beta$t}}$ and $H_{n,n+1}^{\text{FN}}$
are defined in \eqref{eq:Dbetascaled} and \eqref{eq:Dfishnet} respectively.
As discussed in section \ref{sec:eclectic} we call the flavor sequence
$(a_1\cdots a_L)$ \emph{eclectic} if it
 satisfies the hypothesis of the theorem.

The one-loop Hamiltonian density acts as
\begin{equation}
  \mathcal{H}_{n,n+1} |A_n A_{n+1}\rangle = \mathcal{H}^{A'_n A'_{n+1}}_{A_n A_{n+1}} |A'_n A'_{n+1}\rangle.
\end{equation}
The crucial property of the double scaled dilatation operator leading to 
nilpotency is that, whenever $\mathcal{H}^{A'_n A'_{n+1}}_{A_n A_{n+1}} \neq 0$, we have
\begin{equation}
  \flav(A_n) = \flav(A'_{n+1}),\quad \flav(A_{n+1}) = \flav(A'_n),
  \label{eq:chiral-density-one}
\end{equation}
and
\begin{equation}
  \langle \flav(A_n), \flav(A_{n+1}) \rangle \in P_-
  \quad(\text{hence } \langle \flav(A'_n), \flav(A'_{n+1}) \rangle \in P_+).
  \label{eq:chiral-density-two}
\end{equation}
Here $\langle a, b\rangle$ denotes an ordered pair, and we define
\begin{equation}
  P_\pm := \{\langle a, a_{\pm} \rangle | a\in F\} \cup \{\langle a, \bar{a}_\mp \rangle | a\in F\} .
\end{equation}
To prove \eqref{eq:chiral-density-one} and \eqref{eq:chiral-density-two} one uses that $\mathcal{H}$
preserves the R-symmetry charges defined in table \ref{tab:charges}, and 
the presence of the chiral projectors $P^\pm$ in \eqref{eq:Dbetascaled} and \eqref{eq:Dfishnet}.

The strategy of the proof is the following: To each flavor sequence $(a_1  \cdots a_L)$
we are going to assign an integer $d(a_1\cdots a_L)$ which is bounded from above and
such that, for any eclectic sequence,
\begin{equation}
  d(a_1\cdots a_L) < d(a_2 a_1 a_3 \cdots a_L)\, ,
  \label{eq:d-monotonous}
\end{equation}
when $\langle a_1, a_2 \rangle \in P_-$.
By the above remark it is clear that the theorem follows.

An occurrence of a subsequence $(x_1\cdots x_n)$ in $(a_1 \cdots a_L)$ is defined
to be a sequence of indices $\{i_m\}_{m=1,\ldots,n}$ such that
\begin{equation}
  (a_{i_1}a_{i_2}\cdots a_{i_n}) \simeq (x_1 x_2 \cdots x_n), \qquad 1\leq i_1 < i_2 < \cdots < i_n \leq L\, ,
  \label{eq:def-occurrence}
\end{equation}
where $\simeq$ denotes equality modulo cyclic permutations. The multiplicity of $(x_1\cdots x_n)$ in $(a_1 \cdots a_L)$,
denoted $\mul[x_1\cdots x_n;a_1\cdots a_L]$, is the number of occurrences (i.e.~sequences $\{i_m\}_{m=1,\ldots,n}$
such that \eqref{eq:def-occurrence} holds).
We now define $d$ by
\begin{multline}
  d(a_1\cdots a_L) := \sum_{(x_1\cdots x_n)\in C_+} \mul[x_1\cdots x_n;a_1\cdots a_L] \\
   - \sum_{(x_1\cdots x_n)\in C_-} \mul[x_1\cdots x_n;a_1\cdots a_L],
\end{multline}  
where
\begin{equation}
  C_\pm := \{(x_1 x_2\cdots x_n)|n\geq 3, x_i \in F_\noflav, \forall i. \langle x_i, x_{i+1} \rangle \notin P_\mp \}.
\end{equation}
Here and in the remainder we set $x_{n+1} = x_1$.

It remains to show that $d$ satisfies \eqref{eq:d-monotonous}. For the remainder of the
proof we will assume that $\langle a_1, a_2 \rangle \in P_-$.
A little reflection now shows that
\begin{equation}
  \mul[x_1\cdots x_n; a_1\cdots a_L] > \mul[x_1 \cdots x_n; a_2 a_1 a_3 \cdots a_L].
\end{equation}
implies that
there exists an $i\in\{1,\ldots,n\}$ such that $x_i = a_1$ and $x_{i + 1} = a_2$.
But this means that $(x_1 \cdots x_n) \notin C_+$. Similarly, from
\begin{equation}
  \mul[x_1\cdots x_n; a_1\cdots a_L] < \mul[x_1 \cdots x_n; a_2 a_1 a_3 \cdots a_L]
\end{equation}
one can conclude that $(x_1\cdots x_n) \notin C_-$. More loosely, the action of the Hamiltonian
can never decrease the multiplicity of the sequences in $C_+$, and never increase the multiplicity
of the sequences in $C_-$. By the definition of $d$ we thus have
\begin{equation}
  d(a_1\cdots a_L) \leq d(a_2 a_1 a_3 \cdots a_L).
\end{equation}
This holds even when $(a_1\cdots a_L)$ is not eclectic. We show that the inequality is
strict in the eclectic case by a case analysis.

\begin{itemize}
\item Case 1. $a_1 = a_+, a_2 = a$:
Since $(a_1\cdots a_L)$ is eclectic there is an $i \in \{3,\ldots,L\}$ such that
\begin{equation}
  a_i \in \{\bar{a},\bar{a}_+,a_-,\noflav\}.
\end{equation}
Now clearly
\begin{equation}
  \mul[a_2 a_1 a_i; a_1 \cdots a_L] <  \mul[a_2 a_1 a_i; a_2 a_1 a_3 \cdots a_L],
\end{equation}
and also 
\begin{equation}
  (a_2 a_1 a_i) = (a a_+ a_i) \in C_+\, .
\end{equation}
It follows that $d$ increases by at least one. The remaining cases are similar, so
we will be more brief.

\item Case 2. $a_1 = a_-, a_2 = \bar{a}$:
Since $(a_1\cdots a_L)$ is eclectic one of the following three subcases most hold:
\begin{itemize}
\item Subcase 1. 
There is an $i$ such that
\begin{equation}
  a_i \in \{a,\bar{a}_-,\noflav\}.
\end{equation}
This is sufficient since
\begin{equation}
  (a_2 a_1 a_i) = (\bar{a} a_- a_i) \in C_+\, .
\end{equation}
\item Subcase 2. There are $i, j$ such that
\begin{equation}
  a_i= \bar{a}_+\, ,\quad a_j = a_+\, ,\qquad i < j\, .
\end{equation}
This is sufficient since
\begin{equation}
  (a_2 a_1 a_i a_j) = (\bar{a} a_- \bar{a}_+ a_+) \in C_+\, .
\end{equation}
\item Subcase 3. There are $i,j$ such that
\begin{equation}
  a_i= a_+\, ,\quad a_j = \bar{a}_+\, ,\qquad i < j\, .
\end{equation}
This is sufficient since
\begin{equation}
  (a_1 a_2 a_i a_j) = (a_- \bar{a} a_+ \bar{a}_+) \in C_-\, .
\end{equation}
\end{itemize}
\end{itemize}

This concludes the proof. Inspection of the case analysis shows that it would also go through with
$d'(a_1\cdots a_L)$ defined by the same formula as $d$, but with
\begin{equation}
  C'_+ := \{(a a_+ a_-),(a a_+ \bar{a}),(a a_+ \bar{a}_+),(a a_+ \noflav),(a_+\bar{a}\noflav),(\bar{a}a_- \bar{a}_+ a_+) | a\in F\}
\end{equation}
and
\begin{equation}
  C'_- := \{(a_- \bar{a} a_+ \bar{a}_+ | a\in F\}
\end{equation}
instead of $C_\pm$.

\section{\mathversion{bold}Walls in Fishnet Theory and Nilpotency}
\label{app:walls}
\newcommand{\fishletts}{\mathfrak{f}}
Here we will show that the one-loop Hamiltonian of the fishnet model
allows for certain strongly bound states, which we call \emph{walls}.
Furthermore, the Hamiltonian is nilpotent on any state built from $\phi_1$, $\phi_2$ and
derivatives containing such a wall. 

Let $\fishletts_i$, $i=1,2$, denote the set of letters consisting of $\phi_i$ with an arbitrary
number of derivatives, and set $\fishletts = \fishletts_1\cup\fishletts_2$. A wall is a state
on a two-site chain in the subspace
\begin{equation}
  \omega := \Bigl\{ |w\rangle \in \operatorname{span}\bigl\{(A)_1\otimes (B)_2 
                     \bigm| A\in\fishletts_2, B\in\fishletts_1\bigr\} \Bigm| \mathcal{H}_{12} |w\rangle = 0 \Bigr\} .
\end{equation}
It is easy, using Eq.~\eqref{eq:fishnetD2}, to show that $\omega$ contain states with an arbitrary non-zero 
number of derivatives. The simplest example is 
$|(\partial_{1\dot 1} \phi_2) \phi_1\rangle - |\phi_2 (\partial_{1\dot 1} \phi_1)\rangle \in \omega$.

We now consider the space $\mathcal{W}'$ of states on a length $L$ chain with a wall on sites
one and two,
\begin{equation}
  \mathcal{W}' := \operatorname{span} \bigl\{ |w\rangle \otimes |A_3\cdots A_{L}\rangle \bigm| A_i \in \fishletts , \quad |w\rangle \in \omega\bigr\} .
\end{equation}
Note that we are not imposing the zero-momentum constraint for the moment.
Let $\mathcal{H}_{ij} = \mathcal{H}_{ij}^{\text{FN}}$ denote the density operator acting on sites $i$ and $j$ such that
\begin{equation}
  H = \sum_{n=1}^L \mathcal{H}_{n,n+1}\, .
\end{equation}
From the definition of $\omega$ and due to the chiral projector $P^-$ in \eqref{eq:Dfishnet} it is clear 
that $\mathcal{H}_{n,n+1}$ annihilates any state in $\mathcal{W}'$ for $n=L,1,2$.
It follows that
\begin{equation}
  H (|w\rangle \otimes |v\rangle ) = |w\rangle\otimes H_o |v\rangle
\end{equation}
for all $|w\rangle \otimes |v\rangle \in \mathcal{W}'$, where 
\begin{equation}
  H_o :=  \sum_{n=2}^{L-1} \mathcal{H}_{n,n+1} 
\end{equation}
is the Hamiltonian on an \emph{open} spin chain of length $L-2$. By chirality, acting repeatedly with $H_o$ 
will necessarily annihilate any state the latest when the $\phi_2$'s have moved to the right of the $\phi_1$'s. We
conclude that $H$ in nilpotent in $\mathcal{W}'$.

Finally, we need to deal with the zero-momentum constraint. This is easy, we simply project $\mathcal{W}'$ onto the zero-momentum 
subspace. The space of operators containing walls is thus
\begin{equation}
  \mathcal{W} := \bigl\{ P_0|W\rangle \bigm|  |W\rangle \in \mathcal{W'} \bigr\} ,
\end{equation}
with 
\begin{equation}
  P_0 := \frac{1}{L}\sum_{n=0}^{L-1} U^n,
\end{equation}
and where $U$ is the (one-site) translation operator.
Since $P_0$ commutes with the Hamiltonian, we find that $H$ is nilpotent in $\mathcal{W}$.

\bibliography{literature}{}
\bibliographystyle{utphys}

\end{document}